\def\year{2021}\relax
\documentclass[letterpaper]{article} % DO NOT CHANGE THIS
\usepackage{aaai21}  % DO NOT CHANGE THIS
\usepackage{times}  % DO NOT CHANGE THIS
\usepackage{helvet} % DO NOT CHANGE THIS
\usepackage{courier}  % DO NOT CHANGE THIS
\usepackage[hyphens]{url}  % DO NOT CHANGE THIS
\usepackage{graphicx} % DO NOT CHANGE THIS
\usepackage{natbib}  % DO NOT CHANGE THIS AND DO NOT ADD ANY OPTIONS TO IT
\usepackage{caption} % DO NOT CHANGE THIS AND DO NOT ADD ANY OPTIONS TO IT
\usepackage{amsmath}
\usepackage{amsthm}
\usepackage{booktabs}
\usepackage[]{algorithm2e}
\SetKwProg{Fn}{Function}{}{end}\SetKwFunction{FRecurs}{FnRecursive}%

\usepackage[noend]{algpseudocode}
\allowdisplaybreaks

\usepackage{amssymb}
\usepackage{multirow}
\usepackage{tikz}
\usetikzlibrary{calc}

\usepackage{caption}
\usepackage{subcaption}
\usepackage{enumitem}
\DeclareMathOperator{\seq}{\textbf{Seq}}
\DeclareMathOperator{\info}{\textbf{Inf}}

\title{Safe Search for Stackelberg Equilibria in Extensive-Form Games}
\author {
        Chun Kai Ling,\textsuperscript{\rm 1}\footnote{Part of this work was conducted at Facebook AI Research.}
        Noam Brown \textsuperscript{\rm 2}\\
}
\affiliations {
    \textsuperscript{\rm 1} Carnegie Mellon University \\
    \textsuperscript{\rm 2} Facebook AI Research \\
    chunkail@cs.cmu.edu, noambrown@fb.com
}
\begin{document}

\maketitle

\begin{abstract}
    Stackelberg equilibrium is a solution concept in two-player games where the leader has commitment rights over the follower. In recent years, it has become a cornerstone of many security applications, including airport patrolling and wildlife poaching prevention. Even though many of these settings are sequential in nature, existing techniques pre-compute the entire solution ahead of time. In this paper, we present a theoretically sound and empirically effective way to apply search, which leverages extra online computation to improve a solution, to the computation of Stackelberg equilibria in general-sum games. Instead of the leader attempting to solve the full game upfront, an approximate ``blueprint'' solution is first computed offline and is then improved online for the particular subgames encountered in actual play. We prove that our search technique is guaranteed to perform no worse than the pre-computed blueprint strategy, and empirically demonstrate that it enables approximately solving significantly larger games compared to purely offline methods. We also show that our search operation may be cast as a smaller Stackelberg problem, making our method complementary to existing algorithms based on strategy generation.
\end{abstract}

\section{Introduction}
Strong Stackelberg equilibria (SSE) have found many uses in security domains, such as wildlife poaching protection ~\cite{fang2017paws} and airport patrols~\cite{pita2008deployed}. Many of these settings, particularly those involving patrolling, are sequential by nature
%, contain imperfect information and/or chance, and are typically 
and are best
represented as extensive-form games (EFGs). Finding a SSE in general EFGs is provably intractable~\cite{letchford2010computing}. Existing methods convert the problem into a normal-form game and apply column or constraint generation techniques to handle the exponential blowup in the size of the normal-form game~\cite{jain2011quality}. More recent methods cast the problem as a mixed integer linear program (MILP) ~\cite{bosansky2015sequence}. Current state-of-the-art methods build upon this by heuristically generating strategies, and thus avoid considering all possible strategies ~\cite{vcerny2018incremental}.
% However, these methods have few known theoretical guarantees for the size of the final game, and may in theory still require a prohibitive amount of computation. \noam{Is this last sentence adding anything? Seems like it doesn't really matter for this paper.}

All existing approaches for computing SSE are entirely offline. That is, they compute a solution for the entire game ahead of time and always play according to that offline solution. In contrast, \textbf{search} additionally leverages online computation to improve the strategy for the specific situations that come up during play. Search has been a key component for AI in single-agent settings~\cite{lin1965computer,hart1968formal}, perfect-information games~\cite{tesauro1995temporal,campbell2002deep,silver2016mastering,silver2018general}, and zero-sum imperfect-information games~\cite{moravvcik2017deepstack,brown2017superhuman,brown2019superhuman}. In order to apply search to two-player zero-sum imperfect-information games in a way that would not do worse than simply playing an offline strategy, \textbf{safe search} techniques were developed~\cite{burch2014solving,moravcik2016refining,brown2017safe}. Safe search begins with a \textbf{blueprint} strategy that is computed offline. The search algorithm then adds extra constraints to ensure that its solution is no worse than the blueprint (that is, that it approximates an equilibrium at least as closely as the blueprint). However, safe search algorithms have so far only been developed for two-player zero-sum games.

In this paper, we extend safe search to SSE computation in \textit{general-sum} games. We begin with a blueprint strategy for the leader, which is typically some solution (computed offline) of a simpler abstraction of the original game. The leader follows the blueprint strategy for the initial stages of the game, but upon reaching particular subgames of the game tree, computes a refinement of the blueprint strategy \textit{online}, which is then adopted for the rest of the game.

We show that with search, one can approximate SSEs in games much larger than purely offline methods. We also show that our search operation is itself solving a smaller SSE, thus making our method complementary to other methods based on strategy generation. We evaluate our method on a two-stage matrix game, the classic game of Goofspiel, and a larger, general-sum variant of Leduc hold'em. We demonstrate that in large games our search algorithm outperforms offline methods while requiring significantly less computation, and that this improvement increases with the size of the game. Our implementation is publicly available online: \text{https://github.com/lingchunkai/SafeSearchSSE}.

\iffalse
We show that in our application of subgame solving, one can achieve \textit{safety}---that is, assuming a rational follower, applying subgame solving cannot be worse for the leader than applying the blueprint strategy.
%In other words, compared to utilizing the blueprint strategy, our method is `essentially' free, save for
The only cost is
the computation of the blueprint and the refinement of strategies in the subgame encountered, both of which are typically dwarfed by the original problem.
We evaluate our method on the classic game of Goofspiel and a larger, raked variant of Leduc hold'em and show that we are able to outperform offline methods while requiring significantly less computation.
\fi
\section{Background and Related Work} 
%EFGs model sequential interactions between players, and are typically represented as game trees in which each node specifies a state of the game where one player acts (except terminal nodes where no player acts). Chance is considered a distinct player whose strategy is fixed. Non-terminal nodes are partitioned into \emph{information sets}, where the nodes sharing an information set all have the same player acting and are indistinguishable to that player. For example, in poker, nodes which differ only in the private cards of an opponent are indistinguishable to the acting player and therefore share an information set. The terminal nodes of the tree are 2-tuples representing payoffs for each player.

As is standard in game theory, we assume that the strategies of all players, including the algorithms used to compute those strategies, are common knowledge. However, the outcomes of stochastic variables are not known ahead of time.

EFGs model sequential interactions between players, and are typically represented as game trees in which each node specifies a state of the game where one player acts (except terminal nodes where no player acts). In two-player EFGs, there are two players, $\mathcal{P}=\{1,2 \}$. $H$ is the set of all possible nodes $h$ in the game tree, which are represented as sequences of actions (possibly chance). $A(h)$ is the set of actions available in node $h$ and $\mathcal{P}(h) \in \mathcal{P} \cup \{c\}$ is the player acting at that node, where $c$ is the chance player. If a sequence of actions leads from $h$ to $h'$, then we write $h \sqsubset h'$. We denote $Z \subseteq H$ to be the set of all terminal nodes in the game tree. For each terminal node $z$, we associate a payoff for each player, $u_i : Z \rightarrow \mathbb{R}$. For each node $h$, the function $\mathcal{C}: H \rightarrow [0, 1]$ is the probability of reaching $h$, assuming both players play to do so. % Alternatively, $\mathcal{C}(h)$ is the product of probabilities of all chance actions leading to $h$.

Nodes belonging to player $i \in \mathcal{P}$, i.e., $\{h \in H, \mathcal{P}(h) = i \}$ are partitioned into information sets $\mathcal{I}_i$. All nodes $h$ belonging to the same information set $I_i \in \mathcal{I}_i$ are indistinguishable and players must behave the same way for all nodes in $I_i$. Furthermore, all nodes in the same information are required to have the same actions, if $h, h' \in I_i$, then $A(h) = A(h')$. Thus, we overload $A(I_i)$ to define the set of actions in $I_i$. We assume that the game exhibits \textit{perfect recall}, i.e., players do not `forget' past observations or own actions; for each player $i$, the information set $I_i$ is preceded by a unique series of actions and information sets of $i$.
\subsubsection{Sequence Form Representation.}
Strategies in games with perfect recall may be compactly represented in \textit{sequence-form}~\cite{von1996efficient}. A \textit{sequence} $\sigma_i$ is an (ordered) list of actions taken by a single player $i$ in order to reach node $h$. The \textit{empty sequence} $\emptyset$ is the sequence without any actions. The set of all possible sequences achievable by player $i$ is given by $\Sigma_i$. We write $\sigma_i a = \sigma'_i$ if a sequence $\sigma'_i \in \Sigma_i$ may be obtained by appending an action $a$ to $\sigma_i$. With perfect recall, all nodes $h$ in information sets $I_i \in \mathcal{I}_i$ may be reached by a unique sequence $\sigma_i$, which we denote by $\seq_i(I_i)$ or $\seq_i(h)$. Conversely, $\info_i(\sigma_i')$ denotes the information set containing the last action taken in $\sigma_i'$. Using the sequence form, mixed strategies are given by \textit{realization plans}, $r_i: \Sigma_i \rightarrow \mathbb{R}$, which are distributions over sequences. Realization plans for sequences $\sigma_i$ give the probability that this sequence of moves will be played, assuming all other players played such as to reach $\info_i(\sigma_i)$. Mixed strategies obey the sequence form constraints, $\forall i$, $r_{i}(\emptyset)=1 \qquad$, $\forall I_i \in \mathcal{I}_i$, $r_{i}\left(\sigma_{i}\right)=\sum_{a \in A_{i}\left(I_{i}\right)} r_{i}\left(\sigma_{i} a\right)$ and $\sigma_i = \seq_i (I_i)$. 

Sequence forms may be visualized using \textit{treeplexes}~\cite{hoda2010smoothing}, one per player. Informally, a treeplex is a tree rooted at $\emptyset$ with subsequent nodes alternating between information sets and sequences, and are operationally useful for providing recursive implementations for common operations in EFGs such as finding best responses. Since understanding treeplexes is helpful in understanding our method, we provide a brief introduction in the Appendix.

\subsubsection{Stackelberg Equilibria in EFGs.}
Strong Stackelberg Equilibria (SSE) describe games in which there is asymmetry in the commitment powers of players. Here, players $1$ and $2$ play the role of \textit{leader} and \textit{follower}, respectively. The leader is able to commit to a (potentially mixed) strategy and the follower best-responds to this strategy, while breaking ties by favoring the leader. By carefully commiting to a mixed strategy, the leader implicitly issues threats, and followers are made to best-respond in a manner favorable to the leader. SSE are guaranteed to exist and the value of the game for each player is unique. In one-shot games, a polynomial-time algorithm for finding a SSE is given by the multiple-LP approach~\cite{conitzer2006computing}. 

However, solving for SSE in EFGs in general-sum games with either chance or imperfect information is known to be NP-hard in the size of the game tree~\cite{letchford2010computing} due to the combinatorial number of pure strategies. \citet{bosansky2015sequence} avoid transformation to normal form and formulate a compact mixed-integer linear program (MILP) which uses a binary sequence-form follower best response variable to modestly-sized problems. 
% [cite christian gabriele] build upon this to games with uncertain payoffs and with lookahead. 
More recently, \citet{vcerny2018incremental} propose heuristically guided incremental strategy generation. % heuristically guided by the \textit{Stackelberg Extensive Form Correlated Equilibrium}.% to reduce the size of the problem. 

\subsubsection{Safe Search.}
%Unlike past work, our approach targets large games in which strategy generation or heuristics itself may take too long. % , or when heursitic methods do not apply. 
For this paper, we adopt the role of the leader and seek to maximize his expected payoff under the SSE. We assume that the game tree may be broken into several disjoint \textit{subgames}. For this paper, a subgame is defined as a set of states $H_\text{sub} \subseteq H$ such that (a) if $h \sqsubset h'$ and $h \in H_\text{sub}$ then $h' \in H_\text{sub}$, and (b) if $h \in I_i$ and $h \in H_\text{sub}$, then for all $h' \in I_i$, $h' \in H_\text{sub}$. Condition (a) implies that one cannot leave a subgame after entering it, while (b) ensures that information sets are `contained' within subgames---if any history in an information set belongs to a subgame, then every history in that information set belongs to that subgame. For the $j$-th subgame $H^{j}_\text{sub}$, $\mathcal{I}_i^{j} \subseteq \mathcal{I}_i$ is the set of information sets belonging to player $i$ within subgame $j$. Furthermore, let $\mathcal{I}_{i, \text{head}}^{j} \subseteq \mathcal{I}_i^{j}$ be the `head' information sets of player $i$ in subgame $j$, i.e., $I_i \in \mathcal{I}_{i, \text{head}}^{j}$ if and only if $\info_i(\seq_i(I_i))$ does not exist or does not belong to $\mathcal{I}_i^{j}$. With a slight abuse of notation, let $I_{i, head}^j(z)$ be the (unique, if existent) information set in $\mathcal{I}_{i, \text{head}}^{j}$ preceding leaf $z$.

At the beginning, we are given a \textit{blueprint} strategy for the leader, typically the solution of a smaller abstracted game. The leader follows the blueprint strategy in actual play until reaching some subgame. Upon reaching the subgame, the leader computes a refined strategy and follows it thereafter. The pseudocode is given in Algorithm~\ref{alg:generic_sg_solving}.
\begin{algorithm}[t]
\SetKwInOut{Input}{Input}
 \Input{EFG specification, leader blueprint}
 \While{game is not over}{
  \eIf{currently in some subgame $j$}{
   \If{first time in this subgame}{
   (*) Refine leader strategy for subgame $j$
   }
   Play action according to refined strategy 
   }{
   Play action according to blueprint
  }
 }
 \caption{Generic search template.}
 \label{alg:generic_sg_solving}
\end{algorithm}
\iffalse
\begin{algorithm}
        \caption{Generic Search}\label{alg:generic_sg_solving}
        \begin{algorithmic}[1]
            \Procedure{Solve Game}{EFG, blueprint}
            \While{game is not over}
            \If{currently in some subgame $j$}
                \If{first time in subgame $j$}
                \State (*) Refine leader strategy for subgame $j$
                \Else 
                \State Play action according to refined strategy
                \EndIf
            \Else 
                \State Play action according to blueprint
            \EndIf
            \EndWhile
            \EndProcedure
        \end{algorithmic}
    \end{algorithm}
\fi
The goal of the paper is to develop effective algorithms for the refinement step (*). Algorithm~\ref{alg:generic_sg_solving} implicitly defines a leader strategy distinct from the blueprint. Crucially, this implies that the follower responds to this implicit strategy and not the blueprint. %As we will soon see, this results in complications when considering safety.
Search is said to be \textit{safe} when the leader applies Algorithm~\ref{alg:generic_sg_solving}
%such that when the follower best responds, the leader is guaranteed to receive an expected payoff that does no worse than had he played the blueprint (had the follower best responded there as well).
%if the expected value versus the follower's best response is no worse than the expected value of playing the blueprint against the follower's best response to the blueprint.
such that its expected payoff is no less than the blueprint, supposing the follower best responds to the algorithm.
\section{Unsafe Search}
\label{sec:unsafe}
%some of the issues surrounding naive (and unsafe) search.
% to both Algorithm~\ref{alg:generic_sg_solving} and the blueprint.
% This would naturally lead us to our proposed algorithm in later sections. 
To motivate our algorithm, we first explore how unsafe behavior may arise. \textbf{Na\"ive search} assumes that prior to entering a subgame, the follower plays the best-response to the blueprint. For each subgame, the leader computes a normalized distribution of initial (subgame) states and solves a new game with initial states in obeying this distribution. %The parts of the game beneath the subgame are identical to the original game. 
% Alternatively, one can think of this as converting all decision nodes prior to the subgame to chance nodes, where chance acts according to the probabilities of the blueprint and its best-response. 
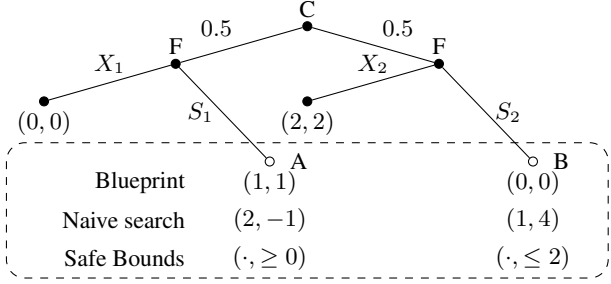
\begin{figure}[t]
\centering
\begin{tikzpicture}[scale=1,font=\footnotesize]
% Two node styles: solid and hollow
\tikzstyle{solid node}=[circle,draw,inner sep=1.2,fill=black];
\tikzstyle{hollow node}=[circle,draw,inner sep=1.2];
% Specify spacing for each level of the tree
\tikzstyle{level 1}=[level distance=5mm,sibling distance=35mm]
\tikzstyle{level 2}=[level distance=5mm,sibling distance=35mm]
\tikzstyle{level 3}=[level distance=2mm,sibling distance=35mm]
% The Tree
\node(0)[solid node]{}
    child{node(1)[solid node]{}
        child{node(3)[solid node]{}
        %child{node[hollow node]{}edge from parent node[left]{$F$}}
        %child{node[hollow node]{}edge from parent node[right]{$G$}}
        edge from parent node[above ]{$X_1$}
        }
        child[level distance=13mm,sibling distance=25mm]{node(5)[hollow node]{}
        %child{node[hollow node]{}edge from parent node[left]{$F$}}
        %child{node[hollow node]{}edge from parent node[right]{$G$}}
        edge from parent node[left]{$S_1$}
        }
        edge from parent node[above left]{$0.5$}
    }
    child{node(2)[solid node]{}
        child{node(4)[solid node]{}
        %child{node[hollow node]{}edge from parent node[left]{$F$}}
        %child{node[hollow node]{}edge from parent node[right]{$G$}}
        edge from parent node[above ]{$X_2$}
        }
        child[level distance=13mm,sibling distance=25mm]{node(6)[hollow node]{}
        %child{node[hollow node]{}edge from parent node[left]{$F$}}
        %child{node[hollow node]{}edge from parent node[right]{$G$}}
        edge from parent node[right]{$S_2$}
        }
        edge from parent node[above right]{$0.5$}
    };
\node[above]at(0){C};
\node[above]at(1){F};
\node[above]at(2){F};
\node[below]at(3){$(0, 0)$};
\node[below]at(4){$(2, 2)$};
\node(100)[below]at(5){$(1, 1)$};
\node[right=0.15cm]at(5){A};
\node[below]at(6){$(0, 0)$};
\node(200)[below=0.5cm]at(5){$(2, -1)$};
\node(300)[below=1.0cm]at(5){$(\cdot, \geq 0)$};
\node[below=0.5cm]at(6){$(1, 4)$};
\node[right=0.15cm]at(6){B};
\node[below=1.0cm]at(6){$(\cdot, \leq 2)$};
\draw[dashed,rounded corners=7]
($(5)+(-3.5,.25)$)rectangle($(6)+(1.0,-1.55)$);
\node[left=1cm]at(100){Blueprint};
\node[left=1cm]at(200){Naive search};
\node[left=1cm]at(300){Safe Bounds};
\end{tikzpicture}
\caption{Unsafe na\"ive search and its game tree. Boxed regions denote subgames. Expected values for each player under (i) the blueprint strategy and its best response and (ii) under na\"ive search is shown in the box, as are bounds guaranteeing no change of follower strategies after refinement.}
\label{fig:subgame_solving_fail_simple}
\end{figure}

Consider the 2-player EFG in Figure~\ref{fig:subgame_solving_fail_simple}, which begins with chance choosing each branch with equal probability. The follower then decides to e\textbf{(X)}it, or \textbf{(S)}tay, where the latter brings the game into a subgame, denoted by the dotted box. Upon reaching A, the follower recieves an expected value (EV) of $1$ when best responding to the blueprint. Upon reaching B, the follower recieves an EV of $0$ when best responding to the blueprint. Thus, under the blueprint strategy, the follower chooses to stay(exit) on the left(right) branches, and the expected payoff per player is $(1.5, 1.5)$. % Naive refinement would require us to solve the subgame shown in Figure~\ref{fig:subgame_transform}. 
\iffalse
\begin{figure}
    \centering
    \input{tikz/tikz_transformed_tree.tex}
    \caption{Transformed game tree of Figure~\ref{fig:subgame_solving_fail_simple} .}
    \label{fig:subgame_transform}
\end{figure}
\fi

\subsubsection{Example 1.} 
Suppose the leader performs na\"ive search in Figure~\ref{fig:subgame_solving_fail_simple}, which improves the leader's EV in A from $1$ to $2$ but reduces the follower's EV in A from $1$ to $-1$. The follower is aware that the leader will perform this search and thus chooses $X_1$ over $S_1$ even before entering A, since exiting gives a payoff of $0$. Conversely, suppose this search improves the leader's EV in B from $0$ to $1$ and also improves the follower's EV from $0$ to $4$. Then the higher post-search payoff in B causes the follower to switch from $X_2$ to $S_2$. These changes cause the leader's EV to drop from $1.5$ to $0.5$. Thus, sticking to the blueprint is preferable to na\"ive search, which means na\"ive search is \textit{unsafe}.\footnote{This counterexample is because of the general-sum nature of this game, and does not occur in zero-sum games.}\\
\textbf{Insight:} Na\"ive search may induce changes in the follower's strategy \textit{before} the subgame, which adjusts the probability of entering each state \textit{within} the subgame. If one could enforce that in the refined subgame, payoffs to the follower in A remain no less than $0$, then the follower would continue to stay, but possibly with leader payoffs greater than the blueprint. %That is, the refined strategy is non-trivially better than the blueprint. 
Similarly, we may avoid entering B  by enforcing that follower payoff in B not exceed $2$. % Since the refinement could be the blueprint strategy itself, this constraint does not lead to infeasibility.
% \paragraph{Example: Solving multiple subgames.} Another problem occurs when there are multiple subgames which are solved separately. This is shown in Figure~\ref{fig:subgame_solving_multiple_fail}, and is similar to the issue suggested by~\cite{brown2017safe}. This example is similar to the last, except for slightly different payoffs, and the addition of a chance node after the follower chooses to stay. There are 2 (identical) subgames, and the outcome of the second chance node determines which subgame is entered. Suppose one performs subgame solving \textit{only} on the left subgame. Then, the expected payoffs for staying are $(1.5, 0)$ for both $S_1$ and $S_2$. Using the tiebreaking rules for SSE, the follower continues to favor staying over exiting. After refinement, the leader expects a payoff (for the full game) of $1.5$, which is greater than $1.0$ of the blueprint.

\subsubsection{Example 2.} %This issue occurs in zero-sum games as well~\citet{brown2017safe}. 
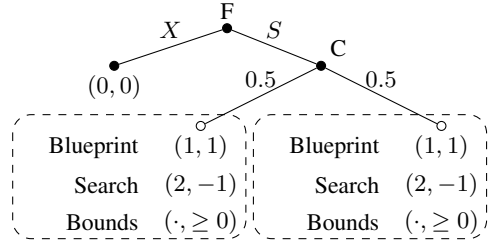
\begin{figure}[t]
    \centering
    \begin{tikzpicture}[scale=1,font=\footnotesize]
% Two node styles: solid and hollow
\tikzstyle{solid node}=[circle,draw,inner sep=1.2,fill=black];
\tikzstyle{hollow node}=[circle,draw,inner sep=1.2];
% Specify spacing for each level of the tree
\tikzstyle{level 1}=[level distance=5mm,sibling distance=30mm]
\tikzstyle{level 2}=[level distance=8mm,sibling distance=32mm]
\tikzstyle{level 3}=[level distance=8mm,sibling distance=32mm]
% The Tree
\node(1)[solid node]{}
    %child{node(1)[solid node]{}
        child{node(3)[solid node]{}
        edge from parent node[above]{$X$}
        }
        child[sibling distance=25mm]{node(5)[solid node]{}
        child{node(11)[hollow node]{}edge from parent node[above]{$0.5$}}
        child{node(12)[hollow node]{}edge from parent node[above]{$0.5$}}
        edge from parent node[above]{$S$}
        };
       % edge from parent node[above left]{$1$}
    %};
%\node[above]at(0){C};
\node[above]at(1){F};
\node[above right]at(5){C};
\node[below]at(3){$(0, 0)$};
\node(100)[below]at(11){$(1, 1)$};
\node[below]at(12){$(1, 1)$};
\node(200)[below=0.5cm]at(11){$(2, -1)$};
\node[below=0.5cm]at(12){$(2, -1)$};
\node(300)[below=1.0cm]at(11){$(\cdot, \geq 0)$};
\node[below=1.0cm]at(12){$(\cdot, \geq 0)$};
\draw[dashed,rounded corners=7]
($(11)+(-2.5,.15)$)rectangle($(11)+(0.6,-1.5)$);
\node[left=0.7cm]at(100){Blueprint};
\node[left=0.7cm]at(200){Search};
\node[left=0.7cm]at(300){Bounds};
\draw[dashed,rounded corners=7]
($(12)+(-2.5,.15)$)rectangle($(12)+(0.6,-1.5)$);
\node[left=-2.5cm]at(100){Blueprint};
\node[left=-2.5cm]at(200){Search};
\node[left=-2.5cm]at(300){Bounds};

%\node(100)[below]at(5){$(1, 1)$};
%\node[below]at(6){$(0, 0)$};
%\node(200)[below=0.5cm]at(5){$(2, -1)$};
%\node[below=0.5cm]at(6){$(0, 0)$};
%\draw[dashed,rounded corners=7]
%($(5)+(-3.5,.25)$)rectangle($(6)+(1.0,-1.2)$);
%\node[left=1cm]at(100){Blueprint};
%\node[left=1cm]at(200){After refinement};
\end{tikzpicture}
    \caption{Failure when searching in multiple subgames.}
    \label{fig:subgame_solving_multiple_fail}
\end{figure}

Consider the game in Figure~\ref{fig:subgame_solving_multiple_fail}.
Here, the follower chooses to exit or stay before the chance node is reached. If the follower chooses stay, then the chance node determines which of two identical subgames is entered. Under the blueprint, the follower receives an EV of $1$ for choosing stay and an EV of $0$ for choosing exit.

Suppose search is performed \textit{only} in the left subgame, which decreases the follower's EV in that subgame from $1$ to $-1$. Then, the expected payoff for staying is $(1.5, 0)$. The follower continues to favor staying (breaking ties in favor of the leader) and the leader's EV increases from $1.0$ to $1.5$. % After refinement, the leader expects a payoff (for the full game) of $1.5$, which is greater than $1.0$ of the blueprint.
%However, one needs to consider the fact that subgame solving would be performed regardless of which subgame was encountered in play. Hence, from the follower's perspective, the leader's strategy would be the refined one in \textit{both} subgames.

Now suppose search is performed on whichever subgame is encountered during play.
Then the follower knows that his EV for staying will be $-1$ regardless of which subgame is reached, and thus will exit. Exiting decreases the leader's payoff to $0$ compared to the blueprint value of $1$, and thus the search is \textit{unsafe}.\footnote{This issue occurs in zero-sum games as well~\cite{brown2017safe}.}
%In other words, conducting search in whichever subgame is encountered is equivalent to this is equivalent to playing according to the search strategy in \textit{all} subgames.
%However, the leader in fact performs search for whichever subgame he encounters. To the follower, the leader's strategy is always refined.
\\ % The problem here is that taken as a single subgame, refinement allows us to reduce follower payoff in each tree by $2$ (since it is modulated by chance), before swapping to exit. However, one needs to account for this potential drop in follower utilities over \textit{all} subgames.
\textbf{Insight:} Performing search using Algorithm~\ref{alg:generic_sg_solving} is equivalent to performing search for \emph{all} subgames. Even if conducting search only in a \textit{single} subgame does not cause a shift in the follower's strategy, the combined effect of applying search to multiple subgames may. Again, one could remedy this by carefully selecting constraints. If we bound the follower post-search EVs for each of the 2 subgames to be $\geq 0$, then we can guarantee that $X$ would never be chosen. Note that this is not the only scheme which ensures safety, e.g., a lower bound of $1$ and $-1$ for the left and right subgame is safe too. 
\section{Safe Search for SSE}
\label{sec:safe}
% \paragraph{Sufficient Conditions for Safety.}
The crux of our method is to modify na\"ive search such that the follower's best response remains the same even when search is applied. This, in turn, can be achieved by enforcing bounds on the follower's EV in any subgame strategies computed via search. Concretely, our search method comprises 3 steps, (i) preprocess the follower's best response to the blueprint and its values, (ii) identify a set of non-trivial safety bounds on follower payoffs\footnote{One could trivially achieve safety by sticking to the blueprint.}, and (iii) solving for the SSE in the subgame reached constrained to respect the bounds computed in (ii).

% This ensures that the distribution of initial states at each subgame is equivalent to that of the blueprint. % and one may find the Stackelberg equilibrium of the subgame by applying \textit{any} SSE-solving algorithm. 
%As demonstrated by the examples, one could achieve invariance in pre-subgame follower best-responses by enforcing bounds on expected follower payoffs for each sequence. The next step is to construct a set of non-trivial constraints that guarantee this invariance.\footnote{One could trivially achieve safety by sticking to the blueprint.}

%In order to elegantly handle imperfect information, it will be convenient to fall back upon the sequence form and treeplex representation. 
\subsubsection{Preprocessing of Blueprint.}Denote the leader's sequence form blueprint strategy as $r_1^\text{bp}$. We will assume that the game is small enough such that the follower's (pure, tiebreaks leader-favored) best response to the blueprint may be computed--- denote it by $r_2^\text{bp}$. We call the set of information sets which, based on $r_2^\text{bp}$ have non-zero probability of being reached the \textit{trunk}, $T \subseteq \mathcal{I}_2: r_2^{\text{bp}} (\seq_2(T)) = 1$. Next, we traverse the follower's treeplex bottom up and compute the payoffs at each information set and sequence (accounting for chance factors $\mathcal{C}(z)$ for each leaf). We term these as best-response values (BRVs) under the blueprint. These are recursively computed for both $\sigma_2 \in \Sigma_2$ and $I_2 \in \mathcal{I}_2$ recursively, (i) $BRV(I_2) = \max_{\sigma_2 \in A(I_2)} BRV(\sigma_2)$, and (ii) $BRV(\sigma_2) = \sum_{I' \in \mathcal{I}_2: \seq_2(I')=\sigma_2} BRV(I') + \sum_{\sigma_1 \in \Sigma_1} r_1(\sigma_1) g_2(\sigma_1, \sigma_2)$, where $g_i(\sigma_i, \sigma_{-i})$ is the expected utility of player $i$ over all nodes reached when executing the sequence pair $(\sigma_i, \sigma_{-i})$, $g_i(\sigma_i, \sigma_{-i}) = \sum_{h \in Z: \sigma_k = \seq_k(h)} u_i(h) \cdot \mathcal{C}(h)$.
 This processing step involves just a single traversal of the game tree.
\iffalse
\begin{align*}
BRV(I_2) &= \max_{\sigma_2 \in A(I_2)} BRV(\sigma_2) \\
BRV(\sigma_2) &= \sum_{I' \in \mathcal{I}_2: \seq_2(I')=\sigma_2} BRV(I') + \sum_{\sigma_1 \in \Sigma_1} r_1(\sigma_1) g_2(\sigma_1, \sigma_2).
\end{align*}
\fi
\subsubsection{Generating Safety Bounds.}
\label{sec:safety_bounds}
% After this preprocessing step based on the blueprint, we are ready to compute the required bounds. 
Loosely speaking, we traverse the follower's treeplex top down while propagating follower payoffs bounds which guarantee that the follower's best response remains $r_2^{\text{bp}}$. This is recursively done until we reach an information set $I$ belonging to some subgame $j$. The EV of $I$ is then required to satisfy its associated bound for future steps of the algorithm.
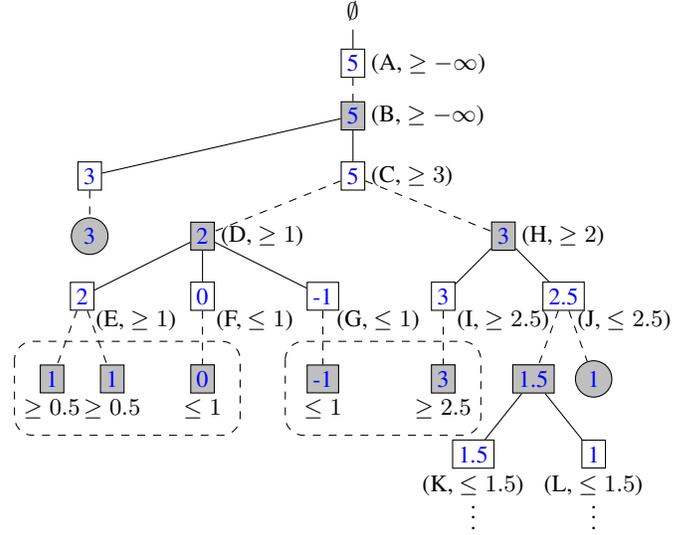
\begin{figure}[t]
    \centering
    \begin{tikzpicture}[scale=1,font=\footnotesize]
% Two node styles: solid and hollow
\tikzstyle{solid node}=[rectangle,draw,inner sep=2.2,fill=lightgray, solid];
\tikzstyle{hollow node}=[rectangle,draw,inner sep=2.2, solid];
\tikzstyle{leaf}=[circle ,draw,inner sep=2.2, solid, fill=lightgray];
% Specify spacing for each level of the tree
\tikzstyle{level 1}=[level distance=7mm,sibling distance=25mm]
\tikzstyle{level 2}=[level distance=7mm,sibling distance=25mm]
\tikzstyle{level 3}=[level distance=8mm,sibling distance=35mm]
\tikzstyle{level 4}=[level distance=8mm,sibling distance=40mm]
\tikzstyle{level 5}=[level distance=8mm,sibling distance=16mm]
\tikzstyle{level 6}=[level distance=11mm,sibling distance=8mm]
\tikzstyle{level 7}=[level distance=10mm,sibling distance=16mm]
% The Tree
\node(0)[right]{$\emptyset$}
    child{node(1)[hollow node]{\color{blue}5} 
        child{node(2)[solid node]{\color{blue}5}
            child{node(3)[hollow node]{\color{blue}3}
                child{node(4)[leaf]{\color{blue}3}edge from parent [dashed] node[right]{}}
                edge from parent [solid] node[left]{}
            }
            child{node(5)[hollow node]{\color{blue}5}
                child{node(6)[solid node]{\color{blue}2}
                    child{node(7)[hollow node]{\color{blue}2}
                        child{node(2111)[solid node]{\color{blue}1}edge from parent [dashed] node[right]{}}
                        child{node(2112)[solid node]{\color{blue}1}edge from parent [dashed] node[right]{}}
                        edge from parent [solid] node[right]{}
                    }
                    child{node(8)[hollow node]{\color{blue}0}
                        child{node(2121)[solid node]{\color{blue}0}edge from parent [dashed] node[right]{}}
                        edge from parent [solid] node[right]{}
                    }
                    child{node(9)[hollow node]{\color{blue}-1}
                        child{node(2131)[solid node]{\color{blue}-1}edge from parent [dashed] node[right]{}}
                        edge from parent [solid] node[right]{}
                    }
                    edge from parent [dashed] node[right]{}
                }
                child{node(10)[solid node]{\color{blue}3}
                    child{node(11)[hollow node]{\color{blue}3}
                        child{node(2211)[solid node]{\color{blue}3}edge from parent [dashed] node[right]{}}
                        edge from parent [solid] node[right]{}
                    }
                    child{node(12)[hollow node]{\color{blue}2.5}
                        child{node(2221)[solid node]{\color{blue}1.5}
                            child{node(22211)[hollow node]{\color{blue}1.5}edge from parent [solid] node[right]{}}
                            child{node(22212)[hollow node]{\color{blue}1}edge from parent [solid] node[right]{}}
                            edge from parent [dashed] node[right]{}
                        }
                        child{node(2222)[leaf]{\color{blue}1}
                            edge from parent [dashed] node[right]{}
                        }
                        edge from parent [solid] node[right]{}
                    }
                    edge from parent [dashed] node[right]{}
                }
                edge from parent [solid] node[right]{}
            }
            child[missing]
            edge from parent [dashed] node[above left]{}
        }
    };
\node[right=0.12cm]at(1){(A, $\geq -\infty$)};
\node[right=0.12cm]at(2){(B, $\geq -\infty$)};
\node[right=0.12cm]at(5){(C, $\geq 3$)};
\node[right=0.12cm]at(6){(D, $\geq 1$)};
\node[below right=0.06cm]at(7){(E, $\geq 1$)};
\node[below right=0.06cm]at(8){(F, $\leq 1$)};
\node[below right=0.06cm]at(9){(G, $\leq 1$)};
\node[right=0.12cm]at(10){(H, $\geq 2$)};
\node[below right=0.06cm]at(11){(I, $\geq 2.5$)};
\node[below right=0.06cm]at(12){(J, $\leq 2.5$)};

\node[below=0.15cm]at(2111){$\geq 0.5$};
\node[below=0.15cm]at(2112){$\geq 0.5$};
\node[below=0.15cm]at(2121){$\leq 1$};
\node[below=0.15cm]at(2131){$\leq 1$};
\node[below=0.15cm]at(2211){$\geq 2.5$};
\node[below=0.15cm]at(22211){(K, $\leq 1.5$)};
\node[below=0.15cm]at(22212){(L, $\leq 1.5$)};
\node[below=0.35cm]at(22211){$\vdots$};
\node[below=0.35cm]at(22212){$\vdots$};
\draw[dashed,rounded corners=7]
($(2111)+(-0.5,.50)$)rectangle($(2121)+(0.5,-0.75)$);
\draw[dashed,rounded corners=7]
($(2131)+(-0.5,.50)$)rectangle($(2211)+(0.5,-0.75)$);
% \draw[dashed,rounded corners=7]
% ($(2221)+(-0.5,.25)$)rectangle($(2222)+(0.5,-0.75)$);
\end{tikzpicture}
    \caption{Example of bounds computation. Filled boxes represent information sets, circled nodes are terminal payoff entries, hollow boxes are sequences which may be followed by parallel information sets, which are in turn preceded by dashed lines. The dashed rectangle indicates subgames, of which we only show the head information sets of. BRVs of sequences and information sets are within the boxes and the (labels, computed bounds) are placed next to them.}
    \label{fig:worked_example_bounds}
\end{figure}
We illustrate the bounds generation process using the worked example in Figure~\ref{fig:worked_example_bounds}. Values of information sets and sequences are in blue and annotated in order of traversal alongside their bounds, whose computation is as follows.
\begin{itemize}[leftmargin=*]
  \setlength{\itemsep}{0pt}
  \setlength{\parskip}{0pt}
    \item The empty sequence $\emptyset$ requires a value greater than $-\infty$.
    \item For each information set (in this case, B) which follows $\emptyset$, we require (vacuously) for their values to be $\geq -\infty$.
    \item We want the sequence C to be chosen. Hence, the value of C has to be $\geq 3$, which, with the lower bound of $-\infty$ gives a final bound of $\geq 3$.
    \item Sum of values for parallel information sets D and H must be greater than C. Under the blueprint, their sum is $5$. This gives a `slack' of $2$, split evenly between D and H, yielding bounds of $2-1=1$ and $3-1=2$ respectively.
    \item Sequence E requires a value no smaller than F, G, and the bound for by the D, which contains it. Other actions have follower payoffs smaller than $1$. We set a \textit{lower} bound of $1$ for E and an \textit{upper} bound of $1$ for F and G.
    \item Sequence I should be chosen over J. Furthermore, the value of sequence I should be $\geq 2$---this was the bound propagated into H. We choose the tighter of the J's blueprint value and the propagated bound of $2$, yielding a bound of $\geq 2.5$ for I and a bound of $\leq 2.5$ for J.
    \item Sequences K and L should not be reached if the follower's best response to the blueprint is followed---we cannot make this portion too appealing. Hence, we apply upper bounds of $1.5$ for sequences K and L.
    % \item Sequences K and L should not be reached if the follower's best response to the blueprint is followed. We need not worry about which action is optimal, except that all sequences satisfy the propagated upper bound (we cannot make this portion of the treeplex too appealing). Hence, we apply upper bounds of $1.5$ for sequences K and L.
\end{itemize}
%A formal description of the bounds generation procedure is given in Algorithm~\ref{alg:bounds_generation} and is deferred to the Appendix. 
A formal description for bounds generation is deferred to the Appendix. The procedure is recursive and identical to the worked example. It takes as input the game, blueprint, best response $r_i^\text{bp}$, and follower BRVs and returns upper and lower bounds $\mathcal{B}(I)$ for all head information sets of subgame $j$, $\mathcal{I}_{2,\text{head}}^j$.
Since the blueprint strategy and its best response satisfies these bounds, feasibility is guaranteed. By construction, lower and upper bounds are obtained for information sets within and outside the trunk respectively. Note also that bounds computation requires only a single traversal of the follower's treeplex, which is smaller than the game tree. 

The bounds generated are not unique. (i) Suppose we are splitting lower bounds at an information set $I$ between child sequences (e.g., the way bounds for sequences E, F, G under information set D were computed). Let $I$ have a lower bound of $\mathcal{B}(I)$ and the best and second best actions $\sigma^*$ and $\sigma'$ under the blueprint is $v^*$ and $v'$ respectively. Our implementation sets lower and upper bounds for $\sigma^*, \sigma'$ to be $\max \left\{ (v^* + v')/2, \mathcal{B}(I) \right\}$. However, any bound of the form $\max \left\{ \alpha \cdot v^* + (1-\alpha) \cdot v', \mathcal{B}(I) \right\}, \alpha \in [0, 1]$ achieves safety. (ii) Splitting lower bounds at sequences $\sigma$ between parallel information sets under $\sigma$ (e.g., when splitting the slack at C between D and H, or in Example 2.). Our implementation splits slack evenly though any non-negative split suffices. We explore these issues in our experiments.
% \subsubsection{MILP Formulation for Subgame Search.}
\subsubsection{MILP formulation for constrained SSE.}
\label{sec:MILP_SS}
Once safety bounds are generated, we can include them in a MILP similar to that of \citet{bosansky2015sequence}. The solution of this MILP is the strategy of the leader, normalized such that $r_{\seq_1(I_1)}$ for all $I_1 \in \mathcal{I}_{1,\text{head}}^j$ is equal to $1$. Let $Z^j$ be the set of terminal states which lie within subgame $j$, $Z^j = Z \cap H_\text{sub}^j$. Let $\mathcal{C}^j(z)$ be the new chance probability when all actions taken prior to the subgame are converted to be by chance, according to the blueprint. That is, $\mathcal{C}^j(z) = \mathcal{C}(z) \cdot  r_1^{\text{bp}} \seq_1(I_{1, \text{head}}(z)) \cdot r_2^{\text{bp}} \seq_2(I_{2, \text{head}}(z))$. Similarly, we set $g^j_2(\sigma_1, \sigma_2) = \sum_{h \in Z^j: \sigma_k = \seq_k(h)} u_2(h) \cdot \mathcal{C}^j(h)$. Let $\mathcal{M}(j)$ be the total probability mass entering subgame $j$ in the original game when the blueprint strategy and best response, $\mathcal{M}(j) = \sum_{z \in Z^j} \mathcal{C}(z) \cdot r_1^{\text{bp}}(\seq_1(z)) \cdot r_2^{\text{bp}}(\seq_2(z))$. 
\begin{align}
    &\max_{p, r, v, s} \sum_{z \in Z^{j}} p(z) u_{1}(z) \mathcal{C}^j(z) \label{eq:mip2_obj}\\
    &v_{\info_{2}\left(\sigma_{2}\right)}=s_{\sigma_{2}}+\sum_{I^{\prime} \in \mathcal{I}_{2} : \seq_{2}\left(I^{\prime}\right)=\sigma_{2}} v_{I^{\prime}}+ \notag \\ 
    & \qquad \qquad \sum_{\sigma_{1} \in \Sigma_{1}} r_{1}\left(\sigma_{1}\right) g^j_{2}\left(\sigma_{1}, \sigma_{2}\right) \qquad \forall \sigma_2 \in \Sigma_2^{j} \label{eq:mip2_slack} \\
    &r_{i}(\sigma_i)=1 \quad \forall i \in \{ 1, 2 \}, I_i \in \mathcal{I}_{i, \text{head}}^j:\seq_i(I_i) = \sigma_i \label{eq:mip2_empty_seq} \\
    &r_{i}\left(\sigma_{i}\right)=\sum_{a \in A_{i}\left(I_{i}\right)} r_{i}\left(\sigma_{i} a\right) \notag \\ 
    & \qquad \qquad \forall i \in \{1, 2\} \quad \forall I_{i} \in \mathcal{I}_{i}^{j}, \sigma_{i}=\seq_{i}\left(I_{i}\right) \label{eq:mip2_flow}\\
    &0 \leq s_{\sigma_{2}} \leq\left(1-r_{2}\left(\sigma_{2}\right)\right) \cdot M \qquad  \forall \sigma_{2} \in \Sigma_{2}^{j} \label{eq:mip2_bigM}\\ 
    &0 \leq p(z) \leq r_{2}\left(\seq_{2}(z)\right) \qquad \forall z \in Z^{j} \label{eq:mip2_max_prob_pl2}\\ 
    &0 \leq p(z) \leq r_{1}\left(\seq_{1}(z)\right) \qquad \forall z \in Z^{j} 
    \label{eq:mip2_max_prob_pl1}
    \\ 
    & \sum_{z \in Z^{j}} p(z) \mathcal{C}^j(z) = \mathcal{M}(j)
    \label{eq:mip2_flow_mass}
    \\ 
    & v_{I_2} \geq \mathcal{B}(I_2) \qquad \forall I_2 \in \mathcal{I}_{2, \text{head}}^{j} \cap T 
    \label{eq:mip2_lower_bounds}
    \\
    & v_{I_2} \leq \mathcal{B}(I_2) \qquad \forall I_2 \in \mathcal{I}_{2, \text{head}}^{j} \cap \overline{T} 
    \label{eq:mip2_upper_bounds}
    \\
    &r_{2}\left(\sigma_{2}\right) \in\{0,1\} \qquad \forall \sigma_{2} \in \Sigma_{2}^{j} 
    \label{eq:mip2_follower_binary}
    \\ 
    &0 \leq r_{1}\left(\sigma_{1}\right) \leq 1 \qquad \forall \sigma_{1} \in \Sigma_{1}^{j}
    \label{eq:mip2_leader_01}
\end{align}
Conceptually, $p(z)$ is %the product of player probabilities to reach leaf $z$, 
such that the probability of reaching $z$ is $p(z) \mathcal{C}(z)$. $r_1$ and $r_2$ are the leader and follower sequence form strategies, $v$ is the value of information set when $r$ is adopted and $s$ is the slack for each sequence. % i.e., the difference of the value of an information set and the value of a particular sequence/action within that information set. 

Objective~\eqref{eq:mip2_obj} is the expected payoff \textit{in the full game} that the leader gets from subgame $j$, %, attenuated by the chance of entering the subgame. 
\eqref{eq:mip2_empty_seq} and \eqref{eq:mip2_flow} are sequence form constraints, %, where head information sets of each subgame are constrained to $1$. 
\eqref{eq:mip2_bigM}, \eqref{eq:mip2_max_prob_pl2}, \eqref{eq:mip2_max_prob_pl1}, \eqref{eq:mip2_follower_binary} and \eqref{eq:mip2_leader_01} ensure the follower is best responding, and \eqref{eq:mip2_flow_mass} ensures that the probability mass entering $j$ is identical to the blueprint. Constraints \eqref{eq:mip2_lower_bounds} and \eqref{eq:mip2_upper_bounds} are bounds previously generated and ensure the follower does not deviate from $r_2^{\text{bp}}$ after refinement. We discuss more details of the MILP in the appendix.
\subsubsection{Safe Search as SSE solutions.}  
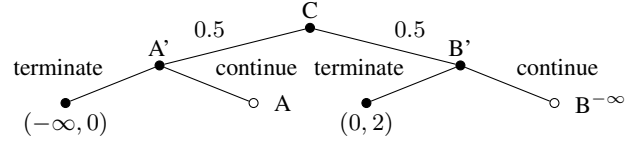
\begin{figure}
% \captionsetup[subfigure]{font=footnotesize}
\centering
\begin{tikzpicture}[scale=1,font=\footnotesize]
% Two node styles: solid and hollow
\tikzstyle{solid node}=[circle,draw,inner sep=1.2,fill=black];
\tikzstyle{hollow node}=[circle,draw,inner sep=1.2];
% Specify spacing for each level of the tree
\tikzstyle{level 1}=[level distance=5mm,sibling distance=40mm]
\tikzstyle{level 2}=[level distance=5mm,sibling distance=25mm]
\tikzstyle{level 3}=[level distance=5mm,sibling distance=40mm]
% The Tree
\node(0)[solid node]{}
    child{node(1)[solid node]{}
        child{node(3)[solid node]{}
        edge from parent node[above left]{terminate}
        }
        child{node(5)[hollow node]{}
        edge from parent node[above right]{continue}
        }
        edge from parent node[above left]{$0.5$}
    }
    child{node(2)[solid node]{}
        child{node(4)[solid node]{}
        edge from parent node[above left]{terminate}
        }
        child{node(6)[hollow node]{}
        edge from parent node[above right]{continue}
        }
        edge from parent node[above right]{$0.5$}
    };
\node[above]at(0){C};
\node[above]at(1){A'};
\node[above]at(2){B'};
\node[below]at(3){$(-\infty, 0)$};
\node[below]at(4){$(0, 2)$};
% \node(100)[below]at(5){$(1, 1)$};
\node[right=0.15cm]at(5){A};
% \node[below]at(6){$(0, 0)$};
\node[right=0.15cm]at(6){$\text{B}^{-\infty}$};
\end{tikzpicture}
%\subcaptionbox{}[.5\textwidth]{
%\input{tikz/tikz_transform_sse.tex}
%}
%\subcaptionbox{}[.5\textwidth]{
%\input{tikz/tikz_after_transform_sse.tex}
%}
%\subcaptionbox{}[.2\textwidth]{%
%\input{tikz_lower_bounds.tex}}
%\subcaptionbox{}[.2\textwidth]{
%\input{tikz_upper_bounds.tex}
%}
\caption{The transformed tree for solving the constrained SSE with the safety bounds of Figure~\ref{fig:subgame_solving_fail_simple}. A' and B' are auxiliary states introduced for the follower. $\text{B}^{-\infty}$ is identical to B, except that leader payoffs are $-\infty$.}
%Enforcing lower bounds via subgame resolving \cite{burch2014solving}. (b) Enforcing upper bounds. The subtree rooted at $h^{-\infty}$ is identical to $h$ except that all subsequent payoffs for the leader are set to $-\infty$. Note that these transformations have to be applied for all states $h \in I_2$.} 
\label{fig:sse_transformation_bounds}
\end{figure}

One is not restricted to using a MILP to enforce these safety bounds. Here we show that the constrained SSE to be solved may be interpreted as the solution to another SSE problem. This implies that we can employ other SSE solvers, such as those involving strategy generation \cite{vcerny2018incremental}. %As such, our method complements other (offline) SSE solvers, potentially allowing us to tackle even larger games. In particular, our method remains viable in games which are too large to even fully specify, as was the case with No-Limit Texas Hold 'em \citet{brown2017superhuman}. As long as computing the best-responses toward the blueprint (with tiebreaking favoring the leader) is computationally feasible, then appropriate bounds \textit{for the subgame which was reached in actual play} may be computed online without having to involve the rest of the game tree.
We briefly describe how transformation is performed on the $j$-th subgame, under the mild assumption that follower head information sets $\mathcal{I}^j_{2, \text{head}}$ are the initial states in $H_\text{sub}^j$. More detail is provided in the Appendix.
%, i.e., if $h \in \mathcal{I}^j_{2, \text{head}}$ then there does not exist $h' \in H_{\text{sub}}^j$ such that $h'\sqsubset h$. We discuss the general case in the Appendix. 
Figure~\ref{fig:sse_transformation_bounds} shows an example construction based on the game in Figure~\ref{fig:subgame_solving_fail_simple}. 
%For each state $h$ in the $j$-th subgame whose predecessor is outside the subgame, we compute the probability of reaching $h$ assuming the follower is playing to reach it (i.e., the product of all leader and chance probabilities along the path to $h$. These probabilities are suitably normalized to $1$; in the transformed game, the root belongs to chance with transitions leading to these states $\{ h \}$. This has the effect of removing parts of the game tree lying outside the subgame.

For every state $h \in \mathcal{I}_{2, \text{head}}^j$, we compute the probability $\omega_h$ of reaching $h$ under the blueprint, assuming the follower plays to reach it. The transformed game begins with chance leading to a normalized distribution of $\omega$ over these states. Now, recall that we need to enforce bounds on follower payoffs for head information sets $I_2 \in \mathcal{I}_{2, \text{head}}^j$. To enforce a lower bound $BRV(I_2) \geq \mathcal{B}(I_2)$, we use a technique similar to \textit{subgame resolving}~\cite{burch2014solving}. Before each state $h \in I_2$, insert an auxiliary state $h'$ belonging to a new information set $I_2'$, where the follower may opt to terminate the game with a payoff of $(-\infty,  \mathcal{B}(I_2)/(\omega_h |I_2|))$ or continue to $h$, whose subsequent states are unchanged.\footnote{The factor $\omega_h |I_2|$ arises since $\mathcal{B}$ was computed in treeplexes, which already takes into account chance.} If the leader's strategy has $BRV(I_2) < \mathcal{B}(I_2)$, the follower would do better by terminating the game, leaving the leader with $-\infty$ payoff.

Enforcing upper bounds $BRV(I_2) \leq \mathcal{B}(I_2)$ may be done analogously. First, we reduce the payoffs to the leader for all leaves underneath $I_2$ to $-\infty$. Second, the follower has an additional action at $I_2$ to terminate the game with a payoff of $\left(0, \mathcal{B}(I_2)/(\omega_h |I_2|) \right)$. If the the follower's response to the leader's strategy gives $BRV(I_2) > \mathcal{B}(I_2)$, then the follower would choose some action other than to terminate the game, which nets the leader $-\infty$. If the bounds are satisfied, then the leader gets a payoff of $0$, which is expected given that an upper bound implies that $I_2$ is not part of the trunk. % This is expected since enforcing an upper bound means that $I_2$ is not part of the trunk and (assuming a rational follower) should never be encountered.
\section{Experiments}
\label{sec:expt}
In this section we show experimental results for our search algorithm (based on the MILP in Section~\ref{sec:MILP_SS}) in synthetic 2-stage games, Goofspiel and Leduc hold'em poker (modified to be general-sum). Experiments were conducted on a Intel i7-7700K @ 4.20GHz with 4 cores and 64GB of RAM. We use the commercial solver Gurobi \cite{gurobi} to solve all instances of MILPs.
%We show that in versions of these games in which computing a Stackelberg equilibrium would be computationally intractable, it is possible to approximate a solution by first computing a blueprint strategy that is a solution to a zero-sum version of the game, and next by applying subgame solving on top of the blueprint strategy to achieve a higher value for the leader. Both of these steps can be done at a far lower computational cost than computing a Stackelberg equilibrium for the entire game.
%As the theory suggests, the leader's value is always higher after applying subgame solving.

We show that even if Stackelberg equilibrium computation for the entire game (using the MILP of ~\citet{bosansky2015sequence}) is warm started using the blueprint strategy $r_1^{\text{bp}}$ and follower's best response $r_2^{\text{bp}}$, then in large games it is still intractable to compute a strategy. In fact, in some cases it is intractable to even generate the model, let alone solve it. In contrast, our safe search algorithm can be done at a far lower computational cost and with far less memory. Since our games are larger than what Gurobi is able to solve to completion in reasonable time, we instead constrain the time allowed to solve each (sub)game and report the incumbent solution. We consider only the time taken by Gurobi in solving the MILP, which dominates preprocessing and bounds generation, both of which only require a constant number of passes over the game tree. In all cases, we warm-start Gurobi with the blueprint strategy. % This results in significant speedups with and without subgame solving. This benefits the latter more, as Gurobi struggles to even find a feasible solution without a warm start.
%In our experiments, we provided a blueprint $r_1^{\text{bp}}$ and follower's best response $r_2^{\text{bp}}$ to warm-start the solver. Not doing so results in a solution significantly worse than the blueprint for both methods (more so for the baseline). We only report the time taken by Gurobi to solve the MILP.
%Here, we compare our algorithm against the baseline method of~\citeauthor{bosansky2015sequence}. Our experiments are conducted on a Intel i7-7700K @ 4.20GHz with 8 cores and 128GB of RAM. We use the commercial solver Gurobi \cite{gurobi} to solve all instances of MILPs. In our experiments, we provided the blueprint $r_1^{\text{bp}}$ and follower's best response $r_2^{\text{bp}}$ to warm-start the solver. Not doing so results in a solution significantly worse than the blueprint for both methods (more so for the baseline). We only report the time taken by Gurobi to solve the MILP. 

To properly evaluate the benefits of search, we perform search on \textit{every} subgame and combine the resulting subgame strategies to obtain the implicit full-game strategy prescribed by Algorithm~\ref{alg:generic_sg_solving}. The follower's best response to this strategy is computed and used to evaluate the leader's payoff. Note that \textit{this is only done to measure how closely the algorithm approximates a SSE}---in practice, search is applied only to the subgame reached in actual play and is performed just once. Hence, the worst-case time for a \textit{single} playthrough is no worse than the longest time required for search over a \textit{single} subgame (and not the sum over all subgames). 

We compare our method against the MILP proposed by \citet{bosansky2015sequence} rather the more recent incremental strategy generation method proposed by \citet{vcerny2018incremental}. The former is flexible and applies to all EFGs with perfect recall, while the latter involves the Stackelberg Extensive Form Correlated Equilibrium (SEFCE) as a subroutine for strategy generation. Computing an SEFCE is itself computationally difficult except in games with no chance, in which case finding an SEFCE can be written as a linear program.

\subsubsection{Two-Stage Games.} The two-stage game closely resembles a 2-step Markov game. In the first stage, both players play a general-sum matrix game $G_{\text{main}}$ of size $n \times n$, after which, actions are made public. In the second stage, one out of $M$ secondary games $\{ G_\text{sec}^{j} \}$, each general-sum and of size $m \times m$ is chosen and played. Each player obtains payoffs equal to the sum of their payoffs for each stage. %The probability of transitioning to game $G_\text{sec}^{j}$ is dependent on the leader's action. 
Given that the leader played action $a_1$, the probability of transitioning to game $j$ is given by the mixture, $\mathbb{P}( G_\text{sec}^{(j)} | a_1 ) = \kappa \cdot X_{j,a_1} + (1-\kappa) \cdot q_j$, where $X_{j,a_1}$ is a $M \times n$ transition matrix non-negative entries and columns summing to $1$ and $q_j$ lies on the $M$ dimensional probability simplex. Here, $\kappa$ governs the level of influence the leader's strategy has on the next stage. % If $\kappa=0$, actions taken in the first stage have no bearing on the second stage. 
\footnote{One may be tempted to first solve the $M$ Stackelberg games independently, and then apply backward induction, solving the first stage with payoffs adjusted for the second. This intuition is incorrect---the leader can issue non-credible threats in the second stage, inducing the follower to behave favorably in the first.}%For instance, the leader may promise `mutual destruction' if the follower does not comply in the first stage.}

The columns of $X$ are chosen by independently drawing weights uniformly from $[0, 1]$ and re-normalizing, while $q$ is uniform. We generate $10$ games each for different settings of $M$, $m$, $n$ and $\kappa$. A subgame was defined for each action pair played in the first stage, together with the secondary game transitioned into. The blueprint was chosen to be the SSE of the first stage \textit{alone}, with actions chosen uniformly at random for the second stage. The SSE for the first stage was solved using the multiple LP method and runs in negligible time ($<5$ seconds). For full-game solving, we allowed Gurobi to run for a maximum of $1000$s. For search, we allowed $100$ seconds---in practice, this never exceeds more than $20$ seconds for \textit{any} subgame.

We report the average quality of solutions in Table~\ref{tbl:synth}. The  full-game solver reports the optimal solution if converges. This occurs in the smaller game settings where ($M=m\leq 10$). In these cases search performs near-optimally. In larger games ($M=m\geq 100$), full-game search fails to converge and barely outperforms the blueprint strategy. In fact, in the largest setting only 2 out of 10 cases resulted in \textit{any} improvement from the blueprint, and even so, still performed worse than our method. Our method yields substantial improvements from the blueprint regardless of $\kappa$.

\begin{table}[]
\centering
\begin{tabular}{ccccccc}
 $n$ & $M$ & $m$ & $\kappa$ &         Blueprint     &   Ours    & Full-game     \\ \hline \hline 
 \multirow{3}{*}{2} & \multirow{3}{*}{2} & \multirow{3}{*}{2} & $0$ & 1.2945 & 1.4472 & \textbf{1.4778} \\
  &  &  & $0.1$ & 1.2945 & 1.4477 & \textbf{1.4779}\\
  &  &  & $0.9$ & 1.2951 & 1.4519 & \textbf{1.4790}\\
  \cline{4-7}
  \multirow{3}{*}{2} & \multirow{3}{*}{10} & \multirow{3}{*}{10} & $0$ & 1.1684 & 1.6179 & \textbf{1.6186} \\
  &  &  & $0.1$ & 1.1689 & 1.6180 & \textbf{1.6186}\\
  &  &  & $0.9$ & 1.1723 & 1.6183 & \textbf{1.6190}\\
  \cline{4-7}
  \multirow{3}{*}{2} & \multirow{3}{*}{100} & \multirow{3}{*}{100} & $0$ & 1.1730 & \textbf{1.6696} & 1.3125 \\
  &  &  & $0.1$ & 1.1729 & \textbf{1.6696} & 1.2652\\
  &  &  & $0.9$ & 1.1722 & \textbf{1.6696} & 1.4055\\
  \cline{4-7}
  \multirow{3}{*}{5} & \multirow{3}{*}{100} & \multirow{3}{*}{100} & $0$ & 1.3756 & \textbf{1.8722} & 1.4074 \\
  &  &  & $0.1$ & 1.3756 & \textbf{1.8723} & 1.4073 \\
  &  &  & $0.9$ & 1.3752 & \textbf{1.8723} & 1.4534\\
  \cline{4-7}
  \hline
\end{tabular}
\caption{Average leader payoffs for two-stage games.}
\label{tbl:synth}
\end{table}

\subsubsection{Goofspiel.}
Goofspiel \cite{ross_1971} is a game where $2$ players simultaneously bid over a sequence of $n$ prizes, valued at $0, \cdots, n-1$. Each player owns cards worth $1, \cdots ,n$, which are used in closed bids for prizes auctioned over a span of $n$ rounds. Bids are public after each round. Cards bid are discarded regardless of the auction outcome. The player with the higher bid wins the prize. In a tie, neither player wins and the prize is discarded. Hence, Goofspiel is not zero-sum, players can benefit by coordinating to avoid ties.

In our setting, the $n$ prizes are ordered uniformly in an order unknown to players. %Players are distinguished and one of them is assumed to be the leader. 
Subgames are selected to be all states which have the same bids and prizes after first $m$ rounds are resolved. 
% Here, $m$ is a parameter controlling the granularity of subgames. 
As $m$ grows, there are fewer but larger subgames. When $m=n$, the only subgame is the entire game. The blueprint was chosen to be the NE under a zero (constant)-sum version of Goofspiel, where players split the prize evenly in ties. The NE of a zero-sum game may be computed efficiently using the sequence form representation \cite{von1996efficient}. Under the blueprint, the leader obtains a utility of  $3.02$ and $5.03$ for $n=4$ and $n=5$ respectively. 

Table~\ref{tbl:goof} summarizes the solution quality and running time as we vary $n, m$. When $n=m$, Gurobi struggles to solve the program to optimality and we report the best incumbent solution found within a shorter time frame. As a sanity check, observe the leader's utility is never worse than the blueprint. When $n=4$, the incumbent solution for solving the full game has improved significantly from the blueprint in fewer than $5$ seconds. This indicates that the game is sufficiently small that performing search is not a good idea. However, when $n=5$, solving the full game ($m=5$) required $180$ times longer compared to search ($m \in \{3, 4\}$) in order to obtain \textit{any} improvement over the blueprint, while search only needed $100$ seconds in order to improve upon the blueprint in a subgame. Furthermore, full-game solving required more than $50$ GB of memory while search required less than $5$ GB. % Furthermore, more than $50$ GB of memory was in full-game solving, while subgame solving with $m=4$ required less than $5$ GB. % These observations demonstrate the massive computational gains from subgame solving.
\iffalse
% larger table too big to fit into 1 col.
\begin{table}
\begin{tabular}{cccccccc}
\multirow{3}{*}{$n$} & 
\multirow{3}{*}{$|\Sigma|$} & 
\multirow{3}{*}{$|\mathcal{I}|$} & 
\multirow{3}{*}{$m$} &
Num. of  &
Num. of &
Time  &
Leader \\
& & & & subgames & binary & allowed per   & utility \\
 & &  & & & variables & subgame (s) &  \\
\hline \hline
  &  &  & 2 & 1728  & 13 & 5 & 3.02   \\
  &  &  & 3 & 64    & 334 & 5 & 3.07  \\ \cline{4-8} 
4 & 2.1 \cdot 10^4 & 1.7 \cdot 10^4 &  &  & &  5 & 4.06  \\
  &  &  & 4 & 1 & 21329 & 100 & 4.15 \\
  &  &  &  &      & & 1000& 4.23 \\
\hline
  &  &  & 3 & 8000  & 334 & 100 & 5.19   \\
  &  &  & 4 & 125    & 21329 & 100 & 5.29  \\ \cline{4-8} 
5 & 2.7 \cdot 10^6 & 2.2 \cdot 10^6 &  &  & &  1000 & 5.03  \\
  &  &  & 5 & 1 & 2666026 & 10000 & 5.03 \\
  &  &  &  &      & & 18000& 5.65 \\
\hline
\end{tabular}
\end{table}
\fi
\begin{table}
\begin{tabular}{cccccc}
\multirow{3}{*}{$n$} & 
\multirow{3}{*}{$(|\Sigma|, |\mathcal{I}|)$} & 
% \multirow{3}{*}{$|\mathcal{I}|$} & 
\multirow{3}{*}{$m$} &
Num. &
Max. time &
\\
& & & of sub- & per sub-  & Leader \\
 & & & games & game (s) &  utility \\
\hline \hline
  & \multirow{5}{*}{$(2.1,1.7) \cdot 10^4$}  & 2 & 1728  &  5 & 3.02   \\
  &   & 3 & 64    &  5 & 3.07  \\ \cline{3-6} 
4 & &  & &  5 & 4.06  \\
  &   & $4^\ddagger$ & 1 &  $1.0 \cdot 10^2$ & 4.15 \\
  &   &  &      & $5.5 \cdot {10^2}^\dagger$ & 4.23 \\
\hline
  & \multirow{5}{*}{$(2.7,2.2) \cdot 10^6$}  & 3 & 8000  & $1.0 \cdot 10^2$ & 5.19   \\
  &   & 4 & 125    & $1.0 \cdot 10^2$ & 5.29  \\ \cline{3-6} 
5 &  &  & &  $1.0 \cdot 10^3$ & 5.03  \\
  &   & $5^\ddagger$ & 1 &  $1.0 \cdot 10^4$ & 5.03 \\
  &   &  &  & $1.8 \cdot {10^4}^\dagger$ & 5.65 \\
\hline
\end{tabular}
\caption{Results for Goofspiel. $^\dagger$This is the earliest time that the incumbent solution achieves the given utility. $^\ddagger$This is equivalent to full-game search.} 
\label{tbl:goof}
\end{table}
\vspace{-0.02in}
\subsubsection{Leduc Hold'em.}
%We experiment on the game of Leduc hold'em~\cite{southey2012bayes}, 
Leduc hold'em~\cite{southey2012bayes} is a simplified form of Texas hold'em. % and is played in a 2-suit deck of cards. 
Players are dealt a single card in the beginning. In our variant there are $n$ cards with 2 suits, 2 betting rounds, an initial bet of $1$ per player, and a maximum of 5 bets per round. The bet sizes for the first and second round are $2$ and $4$. In the second round, a public card is revealed. If a player's card matches the number of the public card, then he/she wins in a showdown, else the higher card wins (a tie is also possible). 

Our variant of Leduc includes \textit{rake}, which is a commission fee to the house. We assume for simplicity a fixed rake $\rho=0.1$. %\footnote{In reality, raked values depend on the size of the pot.} 
This means that the winner receives a payoff of $(1-\rho)x$ instead of $x$. The loser still receives a payoff of $-x$. When $\rho > 0$, the game is not zero-sum. %In our variant, we let there be $5$ bets per round and $n$ cards per suit (total of $2n$ cards). 
Player 1 assumes the role of leader. 
Subgames are defined to be all states with the same public information from the second round onward. The blueprint strategy was obtained using the unraked ($\rho=0$, zero-sum) variant and is solved efficiently using a linear program. We limited the full-game method to a maximum of $5000$ seconds and $200$ seconds \textit{per subgame} for our method.
%\footnote{Preprocessing overheads are negligible to MILP solving.} 
%If the time limit is reached, we use the best solution found so far. Note that in practice, we would \textit{not} have to refine strategies on all subgames, but only those which are encountered in actual play. 
We reiterate that since we perform search only on subgames encountered in actual play, $200$ seconds is an \textit{upper bound}
% \footnote{The weighted (by probability of entering the subgame) average for each subgame would be more accurate.} 
on the time taken for a single playthrough when employing search (some SSE are easier than others to solve). %The full-game method never fully solved our problems in $5000$ seconds.

\begin{table}[]
\begin{tabular}{ccccccc}
 $n$ & $|\Sigma|$ & $|\mathcal{I}|$ &          Blueprint     &   Ours    & Full-game     \\ \hline \hline 
  3    &   5377     &  2016           &          -0.1738  &   -0.1686    & \textbf{-0.1335}     \\
  4    & 9985         & 3744          &         -0.1905      &   \textbf{-0.1862}  &  -0.1882    \\
  5    & 16001      &  6000           &          -0.2028  &   \textbf{-0.2003} & -0.2028      \\
  6    & 23425      &    8784         &          -0.1832  &   \textbf{-0.1780} & -0.1832      \\
  8    & 42497      &   15936         &          -0.1670  &   \textbf{-0.1609} & N/A          \\
  \hline
\end{tabular}
\caption{Leader payoffs for Leduc hold'em with $n$ cards.}
\label{tbl:poker}
\end{table}
The results are summarized in Table~\ref{tbl:poker}. % As expected, our method performs no worse than the blueprint. 
For large games, the full-game method struggles with improving on the blueprint. In fact, when $n=8$ the number of terminal states is so large that the Gurobi model could not be created even after $3$ hours. Even when $n=6$, model construction took an hour---it had near $7\cdot 10^5$ constraints and $4\cdot 10^5$ variables, of which $2.3 \cdot 10^4$ are binary. 
% The large number of constraints is because each terminal state yields two constraints, \eqref{eq:mip1_max_prob_pl2} and \eqref{eq:mip1_max_prob_pl1}. 
Even when the model was successfully built, no progress beyond the blueprint was made. %\footnote{Without the warm start, the full-game method yielded strategies nearly 10 times worse than the blueprint.}
\subsubsection{Varying Bound Generation Parameters.} %Bounds generated may not necessarily be unique (Section~\ref{sec:safe}). 
We now explore how varying $\alpha$ affects solution quality. Furthermore, we experiment with multiplying the slack (see information sets D and H in Section~\ref{sec:safe}) by a constant $\beta \geq 1$. This results in weaker but potentially unsafe bounds. Results on Goofspiel and Leduc are summarized in Figure~\ref{tbl:alphabeta}. We observe that lower values of $\alpha$ yield slightly better performance in Leduc, but did not see any clear trend for Goofspiel. As $\beta$ increases, we observe significant improvements initially. However, when $\beta$ is too large, performance suffers and even becomes unsafe in the case of Leduc. These results suggest that search may be more effective with principled selections of $\alpha$ and $\beta$, which we leave for future work.

\begin{table}[]
\centering
\begin{tabular}{cccccc}
$\alpha$ & $0.1$ & $0.25$& $0.5^{\dagger\dagger}$& $0.75$ & $0.9$ \\ \hline \hline 
% Goofspiel & 5.3173 & 5.33735 & 5.29118& 5.310221&5.302082\\
Goofspiel & 5.32 & 5.34 & 5.29 & 5.31&5.30\\
Leduc & -0.184 & -0.185 & -0.186  & -0.188 & -0.189\\
\hline 
\end{tabular}
\\
\begin{tabular}{cccccc}
$\beta$ & $1^{\dagger\dagger}$ & $2$& $4$ & $8$ & $16$ \\ \hline \hline 
Goofspiel & 5.29 & 5.35 & 5.55 & 5.56 & 5.50 \\
Leduc & -0.186 & -0.182 & -0.178 & -0.212 & -0.212 \\
\hline 
\end{tabular}
\caption{Leader payoffs for varying $\alpha$ and $\beta$. We consider Goofspiel with $n=5, m=4$ and Leduc Hold'em with $n=4$. Time constraints are the same as previous experiments. $^{\dagger\dagger}$These are the default values for $\alpha$ and $\beta$.}
\label{tbl:alphabeta}
\end{table}
\section{Conclusion}
In this paper, we have extended safe search to the realm of SSE in EFGs. We show that safety may be achieved by adding a few straightforward bounds on the value of follower information sets. 
%This ties in neatly with existing methods based upon MILPs, which already contain these values as auxiliary variables. 
We showed it is possible to cast the bounded search problem as another SSE, which makes our approach complementary to other offline methods. Our experimental results on Leduc hold'em demonstrate the ability of our method to scale to large games beyond those which MILPs may solve. Future work includes relaxing constraints on subgames and extension to other equilibrium concepts. %In the future, we would like to broaden the applicability of our method by relaxing restrictions required by subgames as well as extending subgame solving to other equilibrium concepts.

\bibliography{aaai21}
\appendix
\clearpage
\section{Treeplexes} 
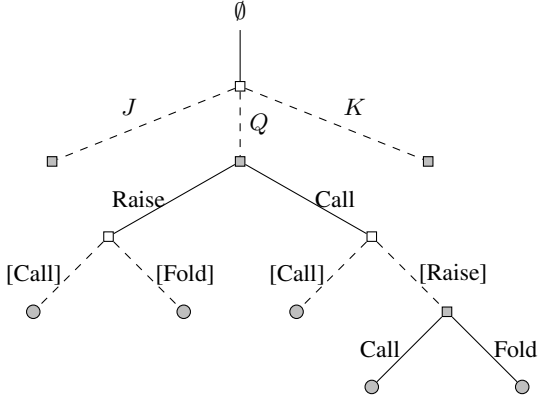
\begin{figure}[ht]
    \centering
    \begin{tikzpicture}[scale=1,font=\footnotesize]
% Two node styles: solid and hollow
\tikzstyle{solid node}=[rectangle,draw,inner sep=1.8,fill=lightgray, solid];
\tikzstyle{hollow node}=[rectangle ,draw,inner sep=1.8, solid];
\tikzstyle{leaf}=[circle ,draw,inner sep=1.8, solid, fill=lightgray];
% Specify spacing for each level of the tree
\tikzstyle{level 1}=[level distance=10mm,sibling distance=25mm]
\tikzstyle{level 2}=[level distance=10mm,sibling distance=25mm]
\tikzstyle{level 3}=[level distance=10mm,sibling distance=35mm]
\tikzstyle{level 4}=[level distance=10mm,sibling distance=20mm]
% The Tree
\node(0)[right]{$\emptyset$}
    child{node(1)[hollow node]{} % empty
        child{node(2)[solid node]{}
        edge from parent [dashed] node[above left]{$J$}
        }
        child{node(2)[solid node]{}
            child{node(21)[hollow node]{}
                child{node[leaf]{}edge from parent [dashed] node[left]{[Call]}}
                child{node[leaf]{}edge from parent [dashed] node[right]{[Fold]}}
            edge from parent [solid] node[left]{Raise}
            }
            child{node(22)[hollow node]{}
                child{node[leaf]{}edge from parent [dashed]  node[left]{[Call]}}
                child{node[solid node]{}
                    child{node[leaf]{}edge from parent [solid] node[left]{Call}}
                    child{node[leaf]{}edge from parent [solid] node[right]{Fold}}
                    edge from parent [dashed]  node[right]{[Raise]}
                }
                edge from parent [solid] node[right]{Call}
            }
            edge from parent [dashed] node[right]{$Q$}
        }
        child{node(3)[solid node]{}
        edge from parent [dashed] node[above right]{$K$}
        }
    };
\end{tikzpicture}
    \caption{Treeplex of player 1 in Kuhn Poker. Filled squares represent information sets, circled nodes are terminal payoffs, hollow squares are points which lead to parallel information sets, which are preceded by dashed lines. Actions/sequences are given by full lines, and information from the second player is in square brackets. The treeplex is `rooted' at the empty sequence. Subtreeplexes for the $J$ and $K$ outcomes are identical and thus omitted.}
    \label{fig:treeplex_pl1_kuhn}
\end{figure}

Algorithms utilizing the sequence form may often be better understood when visualized as \textit{treeplexes}~\cite{hoda2010smoothing,NIPS2018_7366}, with one treeplex defined for each player. Informally, a treeplex may be visualized as a tree with adjacent nodes alternating between information sets and sequences (actions), with the empty sequence forming the root of the treeplex. An example of a treeplex for Kuhn poker~\cite{kuhn1950simplified} is given in Figure~\ref{fig:treeplex_pl1_kuhn}. In this example, a valid (pure) realization plan would be to raise when one obtains a King, or Jack, and when dealt a Queen, call, and follow by folding if the opponent raises thereafter. Mixed strategies in sequence form may be represented by sequence form constraints for each player $i$, $r_{i}(\emptyset)=1 \qquad$  and  $r_{i}\left(\sigma_{i}\right)=\sum_{a \in A_{i}\left(I_{i}\right)} r_{i}\left(\sigma_{i} a\right)$ for each $I_i \in \mathcal{I}_i$ and $\sigma_i = \seq_i (I_i)$. Graphically, this constraint may be visualized as `flow' conservation constraints at infosets, with this flow being duplicated for parallel information sets.

Operations such as best responses have easy interpretations when visualized as treeplexes. When one player's strategy is fixed, the expected values of all leaves may be determined (multiplied by $\mathcal{C}(h)$ and the probability that the other player selects his required sequence). From there, the value of each information set and sequence may be computed via a bottom-up traversal of the treeplex; when parallel information sets are encountered, their values are summed, and when an information set is reached, we select the action with the highest value. After the treeplex is traversed, actions chosen in each information set describe the behavioral strategy of the best response. % In a similar fashion, one may find the best-response of the follower to the attacker, while respecting tiebreaking rules in SSE.

\section{Algorithm for Computing Bounds}
In Section~\ref{sec:safe}, we provided a worked example of how one could compute a set of non-trivial follower bounds which guarantee safety. Algorithm~\ref{alg:bounds_generation} provides an algorithmic description of how this could be done.
\begin{algorithm}
\small
\SetAlgoLined
\SetKwInOut{Input}{Input}
\SetKwInOut{Output}{Output}
\Fn{\textsc{ComputeBounds}} {
     \Input{EFG specification, Blueprint and its BRVs}
     \Output{Bounds $\mathcal{B}(I)$ for all $I \in \mathcal{I}^j_{2, \text{head}}$}
    \textsc{ExpSeqTrunk}($\emptyset, -\infty$)
}
\Fn{\textsc{ExpSeqTrunk}($\sigma$, lb)}{
    \For{$I_2 \in \{ \mathcal{I}_2 | \seq_2(I_2) = \sigma \}$} {
        slack $\leftarrow 
        (\textit{BRV}(\sigma) - lb) / \left| \{ \mathcal{I}_2 | \seq_2(I_2) = \sigma \} \right|$ \\
        \textsc{ExpInfTrunk}($I_2$, $\textit{BRV}(I_2)$-slack) 
    }
}
\Fn{\textsc{ExpInfTrunk}($I$, lb)}{
    \If {$I \in \mathcal{I}^j_{2, \text{head}}$} {
        $\mathcal{B}(I) \leftarrow$ lb\\ 
        \Return 
    }
    $\sigma^*, \sigma' \leftarrow \text{best, second best actions in } I$ under blueprint \\
    $v^*, v' \leftarrow \textit{BRV}(\sigma^*), \textit{BRV}(\sigma')$ \\
    bound $\leftarrow \max \left( \frac{v^*+v'}{2}, lb \right)$ \\
    \For{$\sigma \in \{ \Sigma_2 | \info_2(\sigma) = I \}$} {
        \eIf {$\sigma = \sigma^*$ is in best response}{
            \textsc{ExpSeqTrunk}($\sigma$, bound)
        }{
            \textsc{ExpSeqNonTrunk}($\sigma$, bound)
        }
    }
}
\Fn{\textsc{ExpSeqNonTrunk}($\sigma$, ub)} {
    \For{$I_2 \in \{ \mathcal{I}_2 | \seq_2(I_2) = \sigma \}$} {
        slack $\leftarrow 
        (ub - \textit{BRV}(\sigma)) / \left| \{ \mathcal{I}_2 | \seq_2(I_2) = \sigma \} \right|$ \\
        \textsc{ExpInfNonTrunk}($I_2$, \textit{BRV}($I_2$)+slack) 
    }
}
\Fn{\textsc{ExpInfNonTrunk}($I$, ub)}{
    \If {$I \in \mathcal{I}^j_{2, \text{head}}$} {
        $\mathcal{B}(I) \leftarrow ub$; \Return 
    }
    \For{$\sigma \in \{ \Sigma_2 | \info_2(\sigma) = I \}$} {
        \textsc{ExpSeqNonTrunk}($\sigma, ub$) 
    }
}
\caption{Bounds generation procedure.}
 %\caption{Bounds generation procedure. $\textsc{ExpSeqTrunk}$ and $\textsc{ExpInfTrunk}$ expands sequences and information sets belonging to the trunk, yielding lower bounds. $\textsc{ExpSeqNonTrunk}$ and $\textsc{ExpInfNonTrunk}$ expands sequences and information sets not belonging to the trunk, yielding upper bounds.}
 \label{alg:bounds_generation}
 \end{algorithm}
 
 The \textsc{ComputeBounds} function is the starting point of the bounds generation algorithm. It takes in an EFG specification, a blueprint given, and the BRVs (of sequences $\sigma \in \Sigma_2$ and information sets $I \in \mathcal{I}_2$ computed while preprocessing the blueprint. Our goal is to populate the function $\mathcal{B}(I)$ which maps information sets $I \in \mathcal{I}^j_{2,\text{head}}$ to upper/lower bounds on their values. We begin the recursive procedure by calling \textsc{ExpSeqTrunk} on the empty sequence $\emptyset$ and vacuous lower bound $-\infty$. Note that $\emptyset$ is always in the trunk (by definition).
 
 Specifically, \textsc{ExpSeqTrunk} takes in some sequence $\sigma \in \Sigma_2$ and a lower bound $lb$. Note that we are guaranteed that $lb \leq BRV(\sigma)$. The function compute a set of lower bounds on payoffs of information sets $I_2$ following $\sigma$ such that (a) the best response to the blueprint satisfies these suggested bounds and (b) under the given bounds on values of $I_2$, the follower can at least expect a payoff of $lb$. This is achieved by computing the \textit{slack}, the excess of the blueprint with respect to $lb$ split equally between all $I_2$ following $\sigma$. For each of these $I_2$, we require their value be no smaller than the lower bound given by their BRVs minus the slack. Naturally, this bound is weaker than the BRV itself.
 
 Now, the function \textsc{ExpInfTrunk} does the same bound generation process for a given information set $I$ inside the trunk, given a lower bound $lb$. First, if $I$ is part of the head of a game in subgame $j$, then we simply store $lb$ into $\mathcal{B}(I)$. If not, then we look at all sequences immediately following $I$---specifically, we compare the best and second best sequences, given by $\sigma^*$ and $\sigma'$. To ensure that the best sequence still remains the best response, we need to decide on a threshold \textit{bound} such that (i) all sequences other than $\sigma^*$ does not exceed \textit{bound}, and the value of $\sigma^*$ is no less than \textit{bound} and (ii) the blueprint itself must obey \textit{bound}. One way to specify \textit{bound} is to take the average of the BRVs of $\sigma^*$ and $\sigma'$. For $\sigma^*$ we recursively compute bounds by calling \textsc{ExpSeqTrunk}. For all other sequences, we enter a new recursive procedure which generates \textit{upper bounds}.
 
 The function \textsc{ExpSeqNonTrunk} is similar in implementation to its counterpart \textsc{ExpSeqTrunk}, except that we compute upper instead of lower bounds. Likewise, \textsc{ExpInfoNonTrunk} stores an upper bound if $I$ is in the head of subgame $j$, otherwise, it uses recursive calls to \textsc{ExpInfNonTrunk} to make sure that all immediate sequences following $I$ does not have value greater than $ub$.
 
 In Section~\ref{sec:safe}, we remarked how bounds could be generated in alternative ways, for example, by varying $\alpha$. This would alter the computation of \textit{bound} in \textsc{ExpInfTrunk}. In Section~\ref{sec:expt}, we experiment with increasing the slack by some factor $\beta \geq 1$. That is, we alter the computation of slack in \textsc{ExpSeqTrunk} by multiplying it by $\beta$. Note that this can potentially lead to unsafe behavior, since the follower's payoff under this sequence may possibly be strictly less than $lb$.
 
\section{Details of MILP formulation for SSE}
First, we review the MILP of \citet{bosansky2015sequence}.
\begin{align}
    &\max _{p, r, v, s} \sum_{z \in Z} p(z) u_{1}(z) \mathcal{C}(z) \label{eq:MIP_old_obj}\\
    &v_{\mathrm{inf}_{2}\left(\sigma_{2}\right)}=s_{\sigma_{2}}+\sum_{I^{\prime} \in \mathcal{I}_{2} : \seq_{2}\left(I^{\prime}\right)=\sigma_{2}} v_{I^{\prime}}+ \notag \\ 
    & \qquad \qquad \sum_{\sigma_{1} \in \Sigma_{1}} r_{1}\left(\sigma_{1}\right) g_{2}\left(\sigma_{1}, \sigma_{2}\right) \qquad \forall \sigma_2 \in \Sigma_2
    \label{eq:MIP_old_slack} \\
    &r_{i}(\emptyset)=1 \qquad \forall i \in \{ 1, 2\} 
    \label{eq:MIP_old_empty_seq}\\
    &r_{i}\left(\sigma_{i}\right)=\sum_{a \in A_{i}\left(I_{i}\right)} r_{i}\left(\sigma_{i} a\right)  \notag \\ 
    &\qquad \qquad \forall i \in \{ 1, 2 \} \quad \forall I_{i} \in \mathcal{I}_{i}, \sigma_{i}=\seq_{i}\left(I_{i}\right) 
    \label{eq:MIP_old_seq_form_constr}
    \\
    &0 \leq s_{\sigma_{2}} \leq\left(1-r_{2}\left(\sigma_{2}\right)\right) \cdot 
    M \qquad  \forall \sigma_{2} \in \Sigma_{2} 
    \label{eq:MIP_old_slack_bounds}
    \\ 
    &0 \leq p(z) \leq r_{2}\left(\seq_{2}(z)\right) \qquad \forall z \in Z \label{eq:mip1_max_prob_pl2}\\ 
    &0 \leq p(z) \leq r_{1}\left(\seq_{1}(z)\right) \qquad \forall z \in Z \label{eq:mip1_max_prob_pl1}\\ 
    & \sum_{z \in Z} p(z) \mathcal{C}(z) = 1
    \label{eq:mip1_leaf_prob}
    \\ 
    &r_{2}\left(\sigma_{2}\right) \in\{0,1\} \qquad \forall \sigma_{2} \in \Sigma_{2} 
    \label{eq:mip1_pure_follower}
    \\ 
    &0 \leq r_{1}\left(\sigma_{1}\right) \leq 1 \qquad \forall \sigma_{1} \in \Sigma_{1}
    \label{eq:mip1_nonneg_seq}
\end{align}

Conceptually, $p(z)$ is the product of player probabilities to reach leaf $z$, such that the probability of reaching $z$ is $p(z) \mathcal{C}(z)$. The variables $r_1$ and $r_2$ are the leader and follower strategies in sequence form respectively, while $v$ is the EV of each follower information set when $r_1$ and $r_2$ are adopted. $s$ is the (non-negative) slack for each sequence/action in each information set, i.e., the difference of the value of an information set and the value of a particular sequence/action within that information set. The term $g_i(\sigma_i, \sigma_{-i})$ is the EV of player $i$ over all nodes reached when executing a pair of sequences $(\sigma_i, \sigma_{-i})$, $g_i(\sigma_i, \sigma_{-i}) = \sum_{h \in Z: \sigma_k = \seq_k(h)} u_i(h) \cdot \mathcal{C}(h)$.

Constraint \eqref{eq:MIP_old_slack} ties in the values of the information set $v$ to the slack variables $s$ and payoffs. That is, for every sequence $\sigma_2$ of the follower, the value of its preceding information set is equal to the EV of all information sets $I'$ immediately following $\sigma_2$ (second term) added with the payoffs from all leaf sequences terminating with $\sigma_2$ (third term), compensated by the slack of $\sigma_2$. Constraints \eqref{eq:MIP_old_empty_seq} and \eqref{eq:MIP_old_seq_form_constr} are the sequence form constraints \cite{von1996efficient}. Constraint \eqref{eq:MIP_old_slack_bounds} ensures that, for large enough values of $M$, if the follower's sequence form strategy is $1$ for some sequence, then the slack for that sequence cannot be positive, i.e., the follower must be choosing the best action for himself. Constraints \eqref{eq:mip1_max_prob_pl2}, \eqref{eq:mip1_max_prob_pl1}, and \eqref{eq:mip1_leaf_prob} ensure that $p(z)\mathcal{C}(z)$ is indeed the probability of reaching each leaf.  Constraints \eqref{eq:mip1_pure_follower} and \eqref{eq:mip1_nonneg_seq} enforce that the follower's best response is pure, and that sequence form strategies must lie in $[0, 1]$ for all sequences. The objective \eqref{eq:MIP_old_obj} is the expected utility of the leader, which is linear in $p(z)$.

The MILP we propose for solving the constrained subgame is similar in spirit. Constraints \eqref{eq:mip2_slack}-\eqref{eq:mip2_flow_mass}, \eqref{eq:mip2_follower_binary} and \eqref{eq:mip2_leader_01} are analogous to constraints \eqref{eq:MIP_old_slack}-\eqref{eq:mip1_leaf_prob}, \eqref{eq:mip1_pure_follower}, \eqref{eq:mip1_nonneg_seq} except that thy apply to the subgame $j$ instead of the full game. Similarly, the objective \eqref{eq:mip2_obj} is to maximize the payoffs from within subgame $j$. The key addition is constraint \eqref{eq:mip2_lower_bounds} and \eqref{eq:mip2_upper_bounds}, which are precisely the bounds computed earlier when traversing the treeplex.

\section{Transformation of Safe Search into SSE Solutions}
We provide more details on how the constrained SSE can be cast as another SSE problem. The general idea is loosely related to the \textit{subgame resolving} method of \citet{burch2014solving}, although our method extends to general sum games, and allows for the inclusion of both upper and lower bounds as is needed for our search operation.

The broad idea behind \citet{burch2014solving} is to (i) create an initial chance node leading to all leading states in the subgame (i.e., all states $h \in H_{\text{sub}}^j$ such that there are no states $h' \in H_{\text{sub}}^j$ such that $h' \sqsubset h$) based on the normalized probability of encountering those states under $r_i^{\text{bp}}$ and (ii) enforce the constraints using a \textit{gadget}; specifically, by adding a small number of auxiliary information sets/actions to help coax the solution to obey the required bounds.

\subsection{Restricted case: initial states $h$ in head information sets}
We make the assumption that the $\mathcal{I}_{2, \text{head}}^j$ is a subset of the initial states in subgame $j$. 

\subsubsection{Preliminaries}
For some sequence form strategy pair $r_1, r_2$ for leader and follower respectively, the expected payoff to player $i$ is given by 
$\sum_{z \in Z: \sigma_i = \seq_i (z)} r_1(\sigma_1) \cdot r_2 (\sigma_2) \cdot u_i (z) \cdot \mathcal{C}(z)$, i.e., the summation of the utilities $u_i(z)$ of each leaf of the game, multiplied by the probability that both players play the required sequences $r_i$ and the chance factor $\mathcal{C}(z)$. That is, the utility from each leaf $z$ is weighed by the probability of reaching it $r_1(\sigma_1) \cdot r_2(\sigma_2) \cdot \mathcal{C}(z)$. The value of an information set $I_i \in \mathcal{I}_i$ is the contribution from all leaves under $I_i$, i.e., $V_i(I_i) = \sum_{h \in Z, h' \in I_i, h' \sqsubset h: \sigma_k = \seq_k (h)} r_1(\sigma_1) \cdot r_2(\sigma_2) \cdot u_i(h) \cdot \mathcal{C}(h)$, taking into account the effect of chance for each leaf. 

Now let $b_i(\sigma_i)$ be the behavioral strategy associated with $\sigma_i$, i.e., $b_i(\sigma_i) = r_i (\sigma_i) / r_i (\info(\seq(\sigma_i))) \text{ if } r_i > 0, b_i = 0 \text{ otherwise}. $. The sequence form $r_i(\sigma_i)$ is the product of behavioral strategies in previous information sets. Hence, each of these terms in $V_i$ (be it from leader, follower, or chance) can be separated into products involving those \textit{before} or \textit{after} subgame $j$. That is, for a leaf $z \in Z^j$, the probability of reaching it can be written as $p(z) = \hat{r}^j_1 (z) \cdot  \hat{r}^j_2(z) \cdot  \hat{\mathcal{C}}^j(z) \cdot \check{r}^j_1 (z) \cdot  \check{r}^j_2(z) \cdot  \check{\mathcal{C}}^j(z)$, where $\hat{(\cdot)}^j$ and $\check{(\cdot)}^j$ represent probabilities accrued before and after subgame $j$ respectively. 

\subsubsection{The original game.}
\usetikzlibrary{shapes.geometric}
\usetikzlibrary{decorations.pathmorphing}
\tikzset{
itria/.style={
  draw,dashed,shape border uses incircle,
  isosceles triangle,shape border rotate=90,yshift=-1.45cm},
rtria/.style={
  draw,dashed,shape border uses incircle,
  isosceles triangle,isosceles triangle apex angle=90,
  shape border rotate=-45,yshift=0.2cm,xshift=0.5cm},
ritria/.style={
  draw,dashed,shape border uses incircle,
  isosceles triangle,isosceles triangle apex angle=110,
  shape border rotate=-55,yshift=0.1cm},
letria/.style={
  draw,dashed,shape border uses incircle,
  isosceles triangle,isosceles triangle apex angle=110,
  shape border rotate=235,yshift=0.1cm},
trape/.style={ % requires library shapes.geometric
        draw,
        trapezium,
        trapezium left angle=30,
        trapezium right angle = 30,
        shape border rotate=0,
        text width=3.5cm,
        align=center,
        minimum height=3cm,
        trapezium stretches=true
    },
}

\begin{figure}[t]
\centering
\begin{tikzpicture}[scale=1,font=\footnotesize]
% Two node styles: solid and hollow
\tikzstyle{solid node}=[circle,draw,inner sep=1.2,fill=black];
\tikzstyle{hollow node}=[circle,draw,inner sep=1.2];
\tikzstyle{leaf}=[rectangle,draw,inner sep=1.2];
% Specify spacing for each level of the tree
\tikzstyle{level 1}=[level distance=30mm,sibling distance=15mm]
\tikzstyle{level 2}=[level distance=10mm,sibling distance=6mm]
\tikzstyle{level 3}=[level distance=7.5mm,sibling distance=35mm]
% The Tree
\node(0)[solid node]{}
    child{node(1)[solid node]{} % 
        child[level distance=25mm, sibling distance=12mm]{node(10)[leaf]{}
            edge from parent[decorate, decoration=snake,
            segment length=6mm]
            node [below right]{$\check{r}^j_1 (z^t) \cdot  \check{r}^j_2 (z^t) \cdot  \check{\mathcal{C}}^j(z^t)$}
        }
        child{node(11)[leaf]{}
            edge from parent[decorate, decoration=snake,
            segment length=6mm]
        }
        edge from parent[decorate, decoration=snake,
        segment length = 6mm, thick] node [above left]{$\omega_{h_{z^t}^j} \cdot 1$}
    }
    child{node(2)[solid node]{}
        edge from parent[decorate, decoration=snake,
        segment length=6mm][thick]
        child{node(20)[leaf]{}
            edge from parent[decorate, decoration=snake,
            segment length=6mm]
        }
        child{node(21)[leaf]{}
            edge from parent[decorate, decoration=snake,
            segment length=6mm]
        }
        % edge from parent node[above right]{$0.5 * \sigma_2(S_2)=0.5$}
    }
    child{node(3)[solid node]{}
        edge from parent[decorate, decoration=snake,
        segment length=6mm]
        child{node(30)[leaf]{}
            edge from parent[decorate, decoration=snake,
            segment length=6mm]
        }
        child{node(31)[leaf]{}
            edge from parent[decorate, decoration=snake,
            segment length=6mm]
        }
    }
    child{node(4)[solid node]{}
        child{node(40)[leaf]{}
            edge from parent[decorate, decoration=snake,
            segment length=6mm] 
        }
        child[level distance=25mm, sibling distance=12mm]{node(41)[leaf]{}
            edge from parent[decorate, decoration=snake,
            segment length=6mm]
        }
        edge from parent[decorate, decoration=snake,
        segment length = 6mm] node [above right]{$\omega_{h_{z^{\hat{t}}}^j} \cdot 0$}
    }
    % away nodes from subgmame
    child[level distance=6mm, sibling distance=12mm]{node(987)[leaf]{}
        edge from parent[dashed] node[above right]{away from $j$}
    };
\node[above]at(0){\text{start of game}};
\node[right=0.3cm]at(1){$h_{z^t}^j$};
\node[right=0.15cm]at(4){$h_{z^{\bar{t}}}^j$};
\node(100)[below=0.08cm]at(10){$z^t$};
\node(101)[below=0.08cm]at(41){$z^{\bar{t}}$};
\node[below right=0.0cm]at(100){$p(z^t)=\omega_{h_{z^t}^j} \cdot \check{r}^j_1 (z^t) \cdot  \check{r}^j_2 (z^t) \cdot  \check{\mathcal{C}}^j(z^t) $};
\node[above left=0.10cm]at(1){$I_2^t$};
\node[above left=0.10cm]at(3){$I_2^{\bar{t}}$};
\node[above left=0.60cm]at(1){Subgame $j$};
% \node[trape]at(0){};
    
\draw[dotted]
($(1)+(-0.65,.75)$)rectangle($(2)+(0.5,-0.4)$);
\draw[dotted]
($(3)+(-0.65,.75)$)rectangle($(4)+(0.9,-0.4)$);
% \draw(1000)[dashed, rounded corners=7]
%($(1)+(-2.2,1.35)$)rectangle($(41)+(0.9,-1.5)$);

\iffalse
child{node(3)[solid node]{}
        %child{node[hollow node]{}edge from parent node[left]{$F$}}
        %child{node[hollow node]{}edge from parent node[right]{$G$}}
        edge from parent node[above ]{$X_1$}
        }
        child[level distance=13mm,sibling distance=25mm]{node(5)[hollow node]{}
        %child{node[hollow node]{}edge from parent node[left]{$F$}}
        %child{node[hollow node]{}edge from parent node[right]{$G$}}
        edge from parent node[left]{$S_1$}
        }
        edge from parent node[above left]{$0.5$}
\fi

%\node[below]at(3){$(0, 0)$};
%\node[below]at(4){$(-5, -5)$};
%\node(100)[below]at(5){$(1, 1)$};
%\node[right=0.15cm]at(5){A};
%\node[below]at(6){$(0, 0)$};
%\node(200)[below=0.5cm]at(5){$(2, -1)$};
%\node(300)[below=1.0cm]at(5){$(\cdot, \geq 0)$};
%\node[below=0.5cm]at(6){$(0, 0)$};
%\node[right=0.15cm]at(6){B};
%\node[below=1.0cm]at(6){$(\cdot, \geq -5)$};
%\draw[dashed,rounded corners=7]
%($(5)+(-3.5,.25)$)rectangle($(6)+(1.0,-1.7)$);
%\node[left=1cm]at(100){Blueprint};
%\node[left=1cm]at(200){Naive refinement};
%\node[left=1cm]at(300){Safe Bounds};
\end{tikzpicture}
\caption{Decomposition of probabilities for subgame $j$. Curly lines indicate a series of actions from either player or chance. The dashed box shows subgame $j$, while dotting boxes are head information sets in $\mathcal{I}_{2, \text{head}}^j$. Thick lines belong to states that are below information sets belonging to the trunk. States that do not lead to subgame $j$ are omitted. Note that the subtrees under the $2$ information sets are \textit{not} disjoint, as information sets of the \textit{leader} can span over both subtrees.}
\label{fig:tikz_subgame_tform1}
\end{figure}
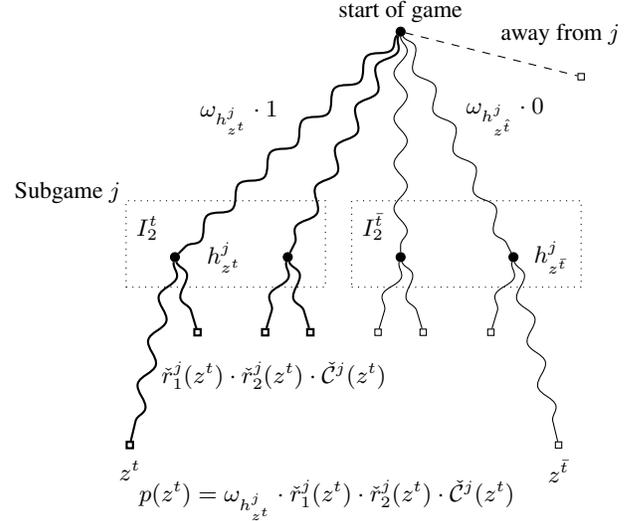
Figure~\ref{fig:tikz_subgame_tform1} illustrates our setting. For head information set $h \in \mathcal{I}_{2, \text{head}}^j$, define $\omega^{\text{bp}}_h$ to be the probability of reaching $h$ following the leader's blueprint assuming the follower plays to reach $h$. Now denote by $h^j_z$ the first state in subgame $j$ leading to leaf $z$ such that $\omega_{h^j_z}^{\text{bp}} = \hat{r}_1^{\text{bp}}(z) \cdot \hat{\mathcal{C}}^j(z)$ is the product of the contributions from the leader and chance, but not the follower. Then, the probability of reaching leaf $z$ is given by $p(z)=\omega_{h^j_z}^{\text{bp}}\cdot \hat{r}_2^{j}(z) \cdot \check{r}^j_1 (z) \cdot  \check{r}^j_2(z) \cdot  \check{\mathcal{C}}^j(z)$. Observe that if $z^t$ lies beneath an infoset $I_2^t \in \mathcal{I}^j_{2, \text{head}} \cap T$ (i.e., it lies in the trunk and the follower under the blueprint plays to $I$), $\hat{r}^j_2(z^t)=1$ (since $r_2^\text{bp}(\seq(I_2))=1$). Conversely, if $z^{\bar{t}}$ lies under $I_2^{\bar{t}} \in \mathcal{I}^j_{2, \text{head}} \cap \bar{T}$, i.e., not part of the trunk, then $\hat{r}^j_2(z^{\bar{t}})=0$, and the probability of reaching the leaf (under $\hat{r}_i^{\text{bp}}$) is $0$. From now onward, we will drop the superscript $(\cdot)^{\text{bp}}$ from $\omega$ when it is clear we are basing it on the blueprint strategy. This is consistent with the notation used in in  Section~\ref{sec:safe}. 

% For each state $I_2^t \in \mathcal{I}^j_{2, \text{head}}$ in the trunk, 
We want to find a strategy $\check{r}_1^j$ such that for every information state $I_2 \in \mathcal{I}^j_{2, \text{head}}$, when $\hat{r}_i=\hat{r}_i^{\text{bp}}$, the best response $\check{r}_2^j$ ensures that $V_2(I_2)$ obeys some upper or lower bounds. That is, value of the information set $V_2$ in this game, given by
\begin{align}
% V_i(I_i) &= \sum_{\substack{z \in Z, h' \in I_i, \\ h' \sqsubset h: \sigma_k = \seq_k (h)}} r_1(\sigma_1) \cdot r_2(\sigma_2) \cdot u_i(z) \cdot \mathcal{C}(z) \\
V_2(I_2) = \sum_{\substack{z \in Z, h \in I_2, \\ h \sqsubset z
% : \sigma_k = \seq_k (h)}} 
}}\omega_{h_z^j} \cdot \check{r}^j_1 (z) \cdot  \check{r}^j_2 (z) \cdot  \check{\mathcal{C}}^j(z)\cdot u_i(z), \label{eq:val_infoset}
\end{align}
should be no greater/less than some $\mathcal{B}(I_2)$.

%Specifically, we have 
%\begin{align*}
%    V_2(I_2) \geq \mathcal{B}(I_2) \qquad &  I_2 \in %\matcal{I}_{2,head}^j \cap T \\
%    V_2(I_2) \leq \mathcal{B}(I_2) \qquad &  I_2 \in %\matcal{I}_{2,head}^j \cap \bar{T}
%\end{align*}

\subsubsection{The transformed game.}
Now consider the transformed subgame described in Section~\ref{sec:safe}. Figure~\ref{fig:tikz_subgame_tform2} illustrates how this transformation may look like and the corresponding probabilities. We look at all possible initial states in subgame $j$, and start the game with chance leading to head states $h$ with a distribution proportional to $\omega_{h}$. For subgame $j$, let the normalizing constant over initial states be $\eta^j > 0$. Note that since we are \textit{including} states outside of the trunk, $\eta^j$ may be greater or less than $1$. % For some leaf $z: \sigma_k = \seq_k(z)$ in the modified game (excluding the auxiliary ones), the probability of reaching it, assuming the auxiliary actions were not taken is $\eta \cdot \omega_h \cdot r_1^{\text{tform}}(\sigma_1) \cdot  r_2^{\text{tform}}(\sigma_2)$, where $r_i^{\text{tform}}$ is the sequence form strategy in the transformed subgame. 
We duplicate every initial state and head information set, giving the follower an option of terminating or continuing on with the game, where terminating yields an immediate payoff of $(-\infty, \frac{\mathcal{B}(I_2^t)}{\omega_{h} |I_2^t|})$ when the information set containing $h$ belongs to the trunk, and $(0, \frac{\mathcal{B}(I_2^{\overline{t}})}{\omega_{h} |I_2^{\overline{t}}|})$ otherwise. For leaves which are descendants of non-trunk information sets, i.e., $z \in Z, h \in I_2 \in \mathcal{I}^j_{2, \text{head}} \cap \bar{T}, h \sqsubset z$, the payoffs for the leaders are adjusted to $-\infty$.
There is a one-to-one correspondence between the behavioral strategies in the modified subgame and the original game simply by using $\check{r}_i^{j}$ interchangeably. 
\usetikzlibrary{shapes.misc}
\tikzset{crossshape/.style={cross out, draw, 
         minimum size=2*(#1-\pgflinewidth), 
         inner sep=0pt, outer sep=0pt}}
\begin{figure}[t]
\centering
\begin{tikzpicture}[scale=1,font=\footnotesize]
% Two node styles: solid and hollow
\tikzstyle{solid node}=[circle,draw,inner sep=1.2,fill=black];
\tikzstyle{hollow node}=[circle,draw,inner sep=1.2];
\tikzstyle{leaf}=[rectangle,draw,inner sep=1.2];
\tikzstyle{cross}=[crossshape,draw,inner sep=3];
% Specify spacing for each level of the tree
\tikzstyle{level 1}=[level distance=35mm,sibling distance=15mm]
\tikzstyle{level 1}=[level distance=25mm,sibling distance=15mm]
\tikzstyle{level 3}=[level distance=10mm,sibling distance=6mm]
\tikzstyle{level 4}=[level distance=7.5mm,sibling distance=35mm]
% The Tree
\node(0)[solid node]{}
    child{node(1)[solid node]{} % 
        child[level distance=9mm, sibling distance=4mm]{node(20001)[leaf]{}
            edge from parent
        }
        child{node(10001)[solid node]{}
            child[level distance=25mm, sibling distance=12mm]{node(10)[leaf]{}
                edge from parent[decorate, decoration=snake,
                segment length=6mm]
                node [below right]{$\check{r}^j_1 (z^t) \cdot  \check{r}^j_2(z^t) \cdot \check{\mathcal{C}}^j(z^t)$}
            }
            child{node(11)[leaf]{}
                edge from parent[decorate, decoration=snake,
                segment length=6mm]
            }
            edge from parent 
        }
        child[level distance=9mm, sibling distance=4mm,missing]{node(20001)[leaf]{}
            edge from parent
        }
        edge from parent node [above left]{$\eta^j \omega_{h^t}$}
    }
    child{node(2)[solid node]{}
        child[level distance=9mm, sibling distance=4mm]{node(20002)[leaf]{}
            edge from parent
        }
        child{node(10002)[solid node]{}
            edge from parent
            child{node(20)[leaf]{}
                edge from parent[decorate, decoration=snake,
                segment length=6mm]
            }
            child{node(21)[leaf]{}
                edge from parent[decorate, decoration=snake,
                segment length=6mm]
            }
            % edge from parent node[above right]{$0.5 * \sigma_2(S_2)=0.5$}
        }
        child[level distance=9mm, sibling distance=4mm,missing]{node(20001)[leaf]{}
            edge from parent
        }
    }
    child{node(3)[solid node]{}
        child[level distance=9mm, sibling distance=4mm,missing]{node(20001)[leaf]{}
            edge from parent
        }
        child{node(10003)[solid node]{}
            edge from parent
            child{node(30)[cross,red]{}
                edge from parent[decorate, decoration=snake,
                segment length=6mm]
            }
            child{node(31)[cross,red]{}
                edge from parent[decorate, decoration=snake,
                segment length=6mm]
            }
        }
        child[level distance=9mm, sibling distance=4mm]{node(20003)[leaf]{}
            edge from parent
        }
    }
    child{node(4)[solid node]{}
        child[level distance=9mm, sibling distance=4mm,missing]{node(20001)[leaf]{}
            edge from parent
        }
        child{node(10004)[solid node]{}
            child{node(40)[cross,red]{}
                edge from parent[decorate, decoration=snake,
                segment length=6mm] 
            }
            child[level distance=25mm, sibling distance=12mm]{node(41)[cross,red]{}
                edge from parent[decorate, decoration=snake,
                segment length=6mm]
            }
            edge from parent
        }
        child[level distance=9mm, sibling distance=4mm]{node(20004)[leaf]{}
            edge from parent
        }
        edge from parent node [above right]{$\eta^j \omega_{h^{\bar{t}}}$}
    };
\node[above]at(0){$C$};
% \node[right=0.3cm]at(10001){$h_z^j$};
% \node[right=0.15cm]at(10004){$h_{z'}^j$};
\node(100)[below=0.08cm]at(10){$z^t$};
\node(101)[below=0.08cm]at(41){$z^{\bar{t}}$};
\node[below right=0.0cm]at(100){$p(z^t)=\eta^j \omega_{h_{z^t}^j} \cdot \check{r}^j_1 (z^t) \cdot  \check{r}^j_2 (z^t) \cdot  \check{\mathcal{C}}^j(z^t) $};
\node[above left=0.10cm]at(10001){$I_2^t$};
\node[above left=0.10cm]at(10003){$I_2^{\bar{t}}$};

\node[above left=0.10cm]at(1){${I_2^t}'$};
\node[above left=0.10cm]at(3){${I_2^{\bar{t}}}'$};

\node[below left=0.00cm]at(20001){$(-\infty, \frac{\mathcal{B}(I_2^t)}{\omega_{h^t} |I_2^t|})$};
% \node[below left=0.00cm]at(20002){$(-\infty, \mathcal{B}(I_2^t))$};
% \node[below right=0.00cm]at(20003){$(0, \mathcal{B}(I_2^{\bar{t}}))$};
\node[below right=0.00cm]at(20004){$(0, \frac{\mathcal{B}(I_2^{\bar{t}})}{\omega_{h^{\bar{t}}} |I_2^{\bar{t}}|})$};

% \node[above left=0.60cm]at(1){Subgame $j$};
% \node[trape]at(0){};
    
\draw[dotted]
($(10001)+(-0.65,.75)$)rectangle($(10002)+(0.5,-0.4)$);
\draw[dotted]
($(10003)+(-0.65,.75)$)rectangle($(10004)+(0.9,-0.4)$);

\draw[dotted]
($(1)+(-0.65,.75)$)rectangle($(2)+(0.5,-0.4)$);
\draw[dotted]
($(3)+(-0.65,.75)$)rectangle($(4)+(0.9,-0.4)$);

%\draw(1000)[dashed, rounded corners=7]
%($(1)+(-2.2,1.35)$)rectangle($(41)+(0.9,-1.5)$);

%\node[below]at(3){$(0, 0)$};
%\node[below]at(4){$(-5, -5)$};
%\node(100)[below]at(5){$(1, 1)$};
%\node[right=0.15cm]at(5){A};
%\node[below]at(6){$(0, 0)$};
%\node(200)[below=0.5cm]at(5){$(2, -1)$};
%\node(300)[below=1.0cm]at(5){$(\cdot, \geq 0)$};
%\node[below=0.5cm]at(6){$(0, 0)$};
%\node[right=0.15cm]at(6){B};
%\node[below=1.0cm]at(6){$(\cdot, \geq -5)$};
%\draw[dashed,rounded corners=7]
%($(5)+(-3.5,.25)$)rectangle($(6)+(1.0,-1.7)$);
%\node[left=1cm]at(100){Blueprint};
%\node[left=1cm]at(200){Naive refinement};
%\node[left=1cm]at(300){Safe Bounds};
\end{tikzpicture}
\caption{An example of the transformed game of Figure~\ref{fig:tikz_subgame_tform1}. Information sets ${I_2^t}'$ and ${I_2^{\bar{t}}}'$ are newly added information sets. Auxiliary actions are in blue, all belonging to newly added information sets. Leaves which are descendants of head information sets not belonging to the trunk are given by red crosses---their payoff to the leader is set to $-\infty$, while keeping the follower payoffs the same.}
\label{fig:tikz_subgame_tform2}
\end{figure}
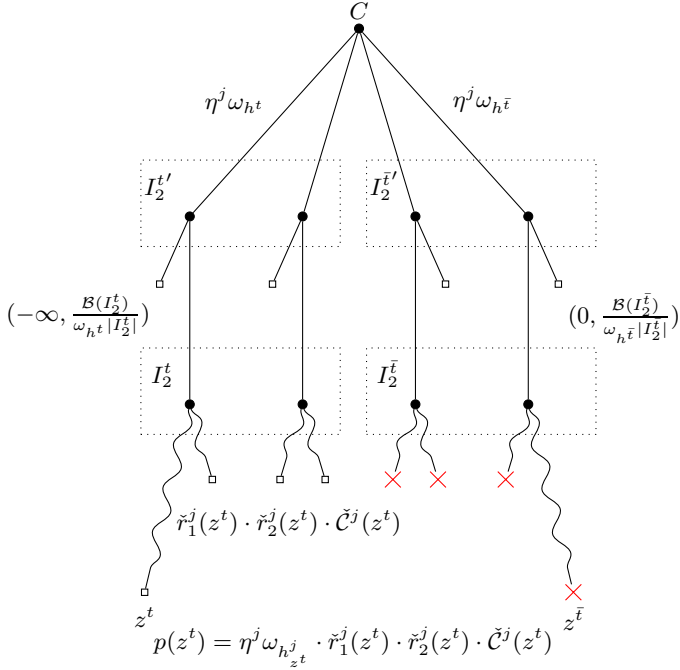
Next, we show that (i) the bounds $\mathcal{B}$ are satisfied by the solution to the transformed game, (ii) for head information sets in the trunk, any solution satisfying $\mathcal{B}$ will never achieve a higher payoff by selecting an auxiliary action, and (iii) for head information sets outside of the trunk, the solutions satisfying $\mathcal{B}$ will, by selecting the auxiliary action, achieve a payoff greater or equal to continuing with the game. For (i), we first consider information set $I_2^{t}$, which is a head infoset also within the trunk. The \textit{terminate} action will result in a follower payoff (taking into account the initial chance node which was added) independent of the leader's subgame strategy $\check{r}_1^j$,
\begin{align}
    \sum_{h \in I_2^{t}} \frac{\mathcal{B}(I_2^t)}{\omega_{h}|I_2^t|} \cdot \eta^j \omega_{h^t} = \eta^j \mathcal{B}(I_2^t).
    \label{eq:eta_bound}
\end{align}
If the follower chooses to continue the game, then his payoff (now dependent on the leader's refined strategy $\check{r}_1^j$ is obtained by performing weighted sums over leaves
\begin{align}
    \eta^j \sum_{\substack{z \in Z, h \in I_2^t \\ h \sqsubset z}} \omega_{h^t} \cdot  \check{r}^j_1 (z) \cdot  \check{r}^j_2 (z) \cdot  \check{\mathcal{C}}^j(z)\cdot u_2(z).
    \label{eq:payoff_stay}
\end{align}
If the leader is to avoid obtaining $-\infty$, then the follower must choose to remain in the game, which will only happen when  \eqref{eq:payoff_stay} $\geq$ \eqref{eq:eta_bound}, i.e.,
\begin{align}
    \sum_{\substack{z \in Z, h \in I_2^t \\ h \sqsubset z}} \omega_{h^t} \cdot  \check{r}^j_1 (z) \cdot  \check{r}^j_2 (z) \cdot  \check{\mathcal{C}}^j(z)\cdot u_2(z) \geq \mathcal{B}(I_2^t).
    \label{eq:bounds_ineq_xform}
\end{align}
The expression on the left hand side of the inequality is precisely the expression in \eqref{eq:val_infoset}. Since the leader can always avoid the $-\infty$ payoff by selecting $\check{r}$ in accordance with the blueprint, the auxiliary action is never chosen and hence, the lower bounds for the value of trunk information sets is always satisfied. 

Similar expressions can be found for non-trunk head-infosets. \eqref{eq:eta_bound} holds completely analogously. The solution to the transformed game needs to make sure the follower always selects the auxiliary action is \textit{always} chosen for information sets not belonging to the trunk, so as to avoid $-\infty$ payoffs from continuing. Therefore, the solution to the transformed game guarantees that \eqref{eq:bounds_ineq_xform} holds, except that the direction of the inequality is reversed. Again, the left hand side of the expression corresponds to \eqref{eq:val_infoset}. Hence, the SSE for the transformed game satisfies our reqired bounds. Furthermore, by starting from \eqref{eq:val_infoset} and working backward, we can also show that any solution $\check{r}_1^j$ satisfying the constrained SSE does not lead to a best response of $-\infty$ for the leader.

Finally, we show that the objective function of the game is identical up to a positive constant. In the original constrained SSE problem, we sum over all leaf descendants of the trunk and compute the leader's utilities weighed by the probability of reaching those leaves.
\begin{align*}
    \sum_{I_2 \in \mathcal{I}^j_{2, \text{head}} \cap T} \sum_{\substack{h \in Z, h' \in I_2\\ h' \sqsubset h}} 
    \omega_{h^j_{z^t}} \cdot \check{r}_1^j(z) \cdot \check{r}_2^j(z) \cdot \mathcal{C}^j(z) \cdot u_1(z)
\end{align*}
the same expression, except for an additional factor of $\eta$. Unlike the constrained SSE setting, the initial distribution has a non-zero probability of starting in a non-trunk state $h \in I_2 \in \mathcal{I}_{2, \text{head}} \cap \bar{T}$. However, since the auxiliary action is always taken under optimality, the leader payoffs from those branches will be $0$.

\subsection{The general case}
The general case is slightly more complicated. Now, the initial states in subgames may not belong to the follower. The issue with trying to add auxiliary states the same way as before is that there could be leader actions lying between the start of the subgame and the (follower) head information set. These leader actions have probabilities which are not yet fixed during the start of the search process. To over come this, instead of enforcing bounds on information sets, we enforce bounds on parts of their parent sequences (which will lie outside the subgame). 

We first partition the head information sets into \textit{groups} based on their parent sequence. Groups can contain singletons. Observe that information sets in the same group are either all in the trunk or all are not. Let the groups be $G_k = \{ I_{2,1}^k, I_{2,2}^k,..., I_{2,m_k}^k\}, I_{2,q}^k \in \mathcal{I}^j_{\text{head}}$, and the group's heads be the initial states which contain a path to some state in the group, i.e, they are:
$G_{k,\text{head}} = \{ h | h \sqsubset h', h' \in I_{2,q}^k \in \mathcal{I}^j_{2, \text{head}} \text{ and } \not \exists h'' \in H^j_\text{sub}, h'' \sqsubset h \}$. Crucially, note that for two distinct groups $G_i, G_k, i\neq k$, their heads $G_{i, \text{head}}$ and $G_{k, \text{head}}$ are disjoint. This because (i) there must be some difference in prior actions that player 2 took (prior to reaching the head information sets) that caused them to be in different groups, and (ii) this action must be taken prior to the subgame by the definition of a head information set.

If 2 information sets $I_{2,1}, I_{2, 2} \in \mathcal{I}_{2, \text{head}}^j$ have the same parent sequence $\sigma_2 = \seq_2 (I_{2,1}) = \seq_2 (I_{2,2})$, i.e., they belong to the same group, $G_k$, it follows that their individual bounds $\mathcal{B}(I_{2,1}), \mathcal{B}(I_{2,2})$ must have come from some split on some bound (upper or lower) on the value of $\sigma_2$. Instead of trying to enforce that the bounds for $I_{2,1}$ and $I_{2,2}$ are satisfied, we try to enforce bounds on the \textit{sum} of the values of $I_{2,1}, I_{2,2}$, since the sum is what is truly important in the bounds for $\sigma_2$ when we perform the bounds generation procedure. 

The transformation then proceeds in the same way as the restricted case, except that we operate on the heads of each \textit{group}, rather than on the head information sets. The bounds for heads of each group is the \textit{sum} of the bounds of head information sets in that group, and that the factor containing the size of the head information set is replaced by the number of heads for that group. 

Upper and lower bounds are enforced using the same gadget as the restricted case, depending on whether the bound is an upper or lower bound. Figure~\ref{fig:tikz_subgame_tform3} shows an example of a lower bound in group $G_k$. Note that the follower payoff in the auxiliary actions contains a sum over bounds over all information sets belonging to the group. Technically, we are performing safe search while respecting weaker (but still safe) bounds. Upper bounds are done the same way analogous to the restricted case. 

\usetikzlibrary{shapes.misc}
\tikzset{crossshape/.style={cross out, draw, 
         minimum size=2*(#1-\pgflinewidth), 
         inner sep=0pt, outer sep=0pt}}
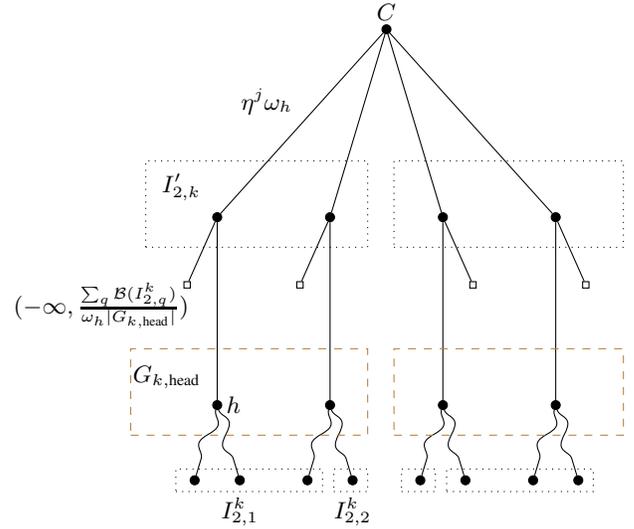
\begin{figure}[ht]
\centering
\begin{tikzpicture}[scale=1,font=\footnotesize]
% Two node styles: solid and hollow
\tikzstyle{solid node}=[circle,draw,inner sep=1.2,fill=black];
\tikzstyle{hollow node}=[circle,draw,inner sep=1.2];
\tikzstyle{leaf}=[rectangle,draw,inner sep=1.2];
\tikzstyle{cross}=[crossshape,draw,inner sep=3];
% Specify spacing for each level of the tree
\tikzstyle{level 1}=[level distance=35mm,sibling distance=15mm]
\tikzstyle{level 1}=[level distance=25mm,sibling distance=15mm]
\tikzstyle{level 3}=[level distance=10mm,sibling distance=6mm]
\tikzstyle{level 4}=[level distance=7.5mm,sibling distance=35mm]
% The Tree
\node(0)[solid node]{}
    child{node(1)[solid node]{} % 
        child[level distance=9mm, sibling distance=4mm]{node(20001)[leaf]{}
            edge from parent
        }
        child{node(10001)[solid node]{}
            child{node(10)[solid node]{}
                edge from parent[decorate, decoration=snake,
                segment length=6mm]
                %node [below right]{$\check{r}^j_1 (z^t) \cdot  \check{r}^j_2(z^t) \cdot \check{\mathcal{C}}^j(z^t)$}
            }
            child{node(11)[solid node]{}
                edge from parent[decorate, decoration=snake,
                segment length=6mm]
            }
            edge from parent 
        }
        child[level distance=9mm, sibling distance=4mm,missing]{node(20001)[leaf]{}
            edge from parent
        }
        edge from parent node [above left]{$\eta^j \omega_{h}$}
    }
    child{node(2)[solid node]{}
        child[level distance=9mm, sibling distance=4mm]{node(20002)[leaf]{}
            edge from parent
        }
        child{node(10002)[solid node]{}
            edge from parent
            child{node(20)[solid node]{}
                edge from parent[decorate, decoration=snake,
                segment length=6mm]
            }
            child{node(21)[solid node]{}
                edge from parent[decorate, decoration=snake,
                segment length=6mm]
            }
            % edge from parent node[above right]{$0.5 * \sigma_2(S_2)=0.5$}
        }
        child[level distance=9mm, sibling distance=4mm,missing]{node(20001)[leaf]{}
            edge from parent
        }
    }
    child{node(3)[solid node]{}
        child[level distance=9mm, sibling distance=4mm,missing]{node(20001)[leaf]{}
            edge from parent
        }
        child{node(10003)[solid node]{}
            edge from parent
            child{node(30)[solid node]{}
                edge from parent[decorate, decoration=snake,
                segment length=6mm]
            }
            child{node(31)[solid node]{}
                edge from parent[decorate, decoration=snake,
                segment length=6mm]
            }
        }
        child[level distance=9mm, sibling distance=4mm]{node(20003)[leaf]{}
            edge from parent
        }
    }
    child{node(4)[solid node]{}
        child[level distance=9mm, sibling distance=4mm,missing]{node(20001)[leaf]{}
            edge from parent
        }
        child{node(10004)[solid node]{}
            child{node(40)[solid node]{}
                edge from parent[decorate, decoration=snake,
                segment length=6mm] 
            }
            child{node(41)[solid node]{}
                edge from parent[decorate, decoration=snake,
                segment length=6mm] 
            }
            edge from parent
        }
        child[level distance=9mm, sibling distance=4mm]{node(20004)[leaf]{}
            edge from parent
        }
        edge from parent node [above right]{}%{$\eta^j \omega_{h}$}
    };
\node[above]at(0){$C$};
% \node[right=0.3cm]at(10001){$h_z^j$};
% \node[right=0.15cm]at(10004){$h_{z'}^j$};
% \node(100)[below=0.08cm]at(10){$z^t$};
% \node(101)[below=0.08cm]at(41){$z^{\bar{t}}$};
% \node[below right=0.0cm]at(100){$p(z^t)=\eta^j \omega_{h_{z^t}^j} \cdot \check{r}^j_1 (z^t) \cdot  \check{r}^j_2 (z^t) \cdot  \check{\mathcal{C}}^j(z^t) $};
\node[above left=0.10cm]at(10001){${G_{k,\text{head}}}$};
% \node[above left=0.10cm]at(10003){$I_2^{\bar{t}}$};

\node[above left=0.10cm]at(1){${I'_{2,k}}$};
% \node[above left=0.10cm]at(3){${I_2^{\bar{t}}}'$};

\node[below=0.10cm]at(11){$I_{2,1}^k$};
\node[below=0.10cm]at(21){$I_{2,2}^k$};

\node[right]at(10001){$h$};

\node[below left=-0.15cm]at(20001){$(-\infty, \frac{\sum_{q}\mathcal{B}(I_{2,q}^k)}{\omega_{h} |G_{k, \text{head}}|})$};
% \node[below left=0.00cm]at(20002){$(-\infty, \mathcal{B}(I_2^t))$};
% \node[below right=0.00cm]at(20003){$(0, \mathcal{B}(I_2^{\bar{t}}))$};
%\node[below right=-0.15cm]at(20004){$(0, \frac{\mathcal{B}(I_2^{\bar{t}}))}{\omega_{h^{\bar{t}}} |I_2^{\bar{t}}|}$};

% \node[above left=0.60cm]at(1){Subgame $j$};
% \node[trape]at(0){};

\draw[dashed,brown]
($(10001)+(-1.15,.75)$)rectangle($(10002)+(0.5,-0.4)$);
\draw[dashed,brown]
($(10003)+(-0.65,.75)$)rectangle($(10004)+(0.9,-0.4)$);

% the tiny guys ***************************
\draw[dotted]
($(10)+(-0.25,.15)$)rectangle($(20)+(0.2,-0.15)$);
\draw[dotted]
($(21)+(-0.25,.15)$)rectangle($(21)+(0.2,-0.15)$);

\draw[dotted]
($(30)+(-0.25,.15)$)rectangle($(30)+(0.2,-0.15)$);
\draw[dotted]
($(31)+(-0.25,.15)$)rectangle($(41)+(0.2,-0.15)$);
% the tiny guys ***************************

\draw[dotted]
($(1)+(-0.95,.75)$)rectangle($(2)+(0.5,-0.4)$);
\draw[dotted]
($(3)+(-0.65,.75)$)rectangle($(4)+(0.9,-0.4)$);

%\draw(1000)[dashed, rounded corners=7]
%($(1)+(-2.2,1.35)$)rectangle($(41)+(0.9,-1.5)$);

%\node[below]at(3){$(0, 0)$};
%\node[below]at(4){$(-5, -5)$};
%\node(100)[below]at(5){$(1, 1)$};
%\node[right=0.15cm]at(5){A};
%\node[below]at(6){$(0, 0)$};
%\node(200)[below=0.5cm]at(5){$(2, -1)$};
%\node(300)[below=1.0cm]at(5){$(\cdot, \geq 0)$};
%\node[below=0.5cm]at(6){$(0, 0)$};
%\node[right=0.15cm]at(6){B};
%\node[below=1.0cm]at(6){$(\cdot, \geq -5)$};
%\draw[dashed,rounded corners=7]
%($(5)+(-3.5,.25)$)rectangle($(6)+(1.0,-1.7)$);
%\node[left=1cm]at(100){Blueprint};
%\node[left=1cm]at(200){Naive refinement};
%\node[left=1cm]at(300){Safe Bounds};
\end{tikzpicture}
\caption{An example of a general transformation. Brown dashed lines are the heads of individual groups. $I_{2,1}^k$ and $I_{2,2}^k$ belong to the same group $G_k$ with the heads $G_{k, \text{head}}$. The newly created auxiliary states are in the new information set $I_{2,k}'$. In this case, $G_k$ is in the trunk, hence we enforce a lower bound being enforced for $G_k$.}
\label{fig:tikz_subgame_tform3}
\end{figure}

\end{document}